\documentclass[]{jfm}
\usepackage{natbib}
\usepackage{amsmath}
\usepackage{amssymb}
\usepackage[dvips]{graphicx}
\usepackage{psfrag}
\usepackage{color}
\usepackage{epstopdf}

\newcommand{\aver}[1]{\left\langle {#1} \right\rangle}
\newcommand{\R}{\mathcal{R}}
\newcommand{\SAV}{\mathcal{S}}
\newcommand{\G}{\mathcal{G}}

\begin{document}

\title[Reynolds-dependence of skin-friction drag reduction by spanwise forcing]
{Reynolds-dependence \\ of turbulent skin-friction drag reduction \\ induced by spanwise forcing}
\author[D. Gatti \& M. Quadrio]%
{
\ns D\ls A\ls V\ls I\ls D\ls E\ns G\ls A\ls T\ls T\ls I$^1$
\and 
\ns M\ls A\ls U\ls R\ls I\ls Z\ls  I\ls O\ns Q\ls U\ls A\ls D\ls R\ls  I\ls O$^2$
\thanks{Email address for correspondence: maurizio.quadrio@polimi.it}
}

\affiliation{$^1$Institute of Fluid Mechanics, Karlsruhe Institute of Technology,
Kaiserstra\ss e 10, 76131 Karlsruhe, Germany
\\[\affilskip]
$^2$Department of Aerospace Science and Technology, Politecnico di Milano, 
via La Masa 34, 20156 Milano, Italy
}

\date{\today}

\maketitle

\begin{abstract}
This paper examines how increasing the value of the Reynolds number $Re$ affects the ability of spanwise-forcing techniques to yield turbulent skin-friction drag reduction. The considered forcing is based on the streamwise-travelling waves of spanwise wall velocity (Quadrio {\em et al. J. Fluid Mech.}, vol. 627, 2009, pp. 161--178). The study builds upon an extensive drag-reduction database created with Direct Numerical Simulation of a turbulent channel flow for two, 5-fold separated values of $Re$, namely $Re_\tau=200$ and $Re_\tau=1000$. The sheer size of the database, which for the first time systematically addresses the amplitude of the forcing, allows a comprehensive view of the drag-reducing characteristics of the travelling waves, and enables a detailed description of the changes occurring when $Re$ increases. The effect of using a viscous scaling based on the friction velocity of either the non-controlled flow or the drag-reduced flow is described.
In analogy with other wall-based drag reduction techniques, like for example riblets, the performance of the travelling waves is well described by a vertical shift of the logarithmic portion of the mean streamwise velocity profile. Except when $Re$ is very low, this shift remains constant with $Re$, at odds with the percentage reduction of the friction coefficient, which is known to present a mild, logarithmic decline. Our new data agree with the available literature, which is however mostly based on low-$Re$ information and hence predicts a quick drop of maximum drag reduction with $Re$. The present study supports a more optimistic scenario, where for an airplane at flight Reynolds numbers a drag reduction of nearly 30\% would still be possible thanks to the travelling waves.
\end{abstract}

\section{Introduction}\label{sec:introduction}

Reducing the turbulent drag in general, and the turbulent skin-friction drag in particular, is a potentially rewarding technological goal which, however, presents several challenges that span from the understanding of a complex physics to the design of a reliable and affordable control system.

In the last few decades, fundamental research efforts in skin-friction drag reduction met with considerable success, and  several viable strategies to reduce drag have been introduced, although often only proofs-of-concept based on numerical simulations or laboratory experiments are available. For obvious reasons, such studies are typically limited to low-Reynolds number flows, and the question naturally arises how to extrapolate the observed performance (in terms of e.g. maximum drag reduction, or energy cost of the control technique) to the higher values of the Reynolds number $Re$ typical of most industrial applications. 

Some of the earliest techniques, however, have already been demonstrated in the envisaged application: the most notable example is perhaps riblets \citep{walsh-1980, bechert-bartenwerfer-1989, garcia-jimenez-2011-a}, i.e. small streamwise-aligned grooves patterned on an otherwise smooth surface, which have been tested on a full-scale airplane in flight conditions \citep{walsh-sellers-mcginley-1989}. Although in flight tests an indirect measure of the riblets effectiveness based on changes in fuel consumption is often preferred, in simple and well-controlled laboratory flows like a channel flow the friction drag reduction achieved by riblets is typically characterized directly in terms of the drag reduction rate $\R$, defined as the relative change of skin-friction coefficient $C_f$ between the controlled and the reference flow: 
\begin{equation}
  \R = 1-\frac{C_f}{C_{f,0}} .
  \label{eq:DRR}
\end{equation}

In this definition, the subscript "0" indicates a quantity measured in the reference flow, and the skin-friction coefficient is defined as 
\begin{equation}
C_f=2 \frac{\tau_w}{\rho U_b^2} ,
\label{eq:Cf}
\end{equation}
where $\tau_w$ is the wall-shear stess, $\rho$ is the fluid density and $U_b$ the bulk velocity. The maximum drag reduction for riblets is known \citep{luchini-1996, spalart-mclean-2011} to vary with $Re$, owing to the (mild) $Re$-dependency of $C_f$ itself. A $Re$-independent quantification of the riblets ability to reduce drag is obtained by considering riblets as a particular roughness distribution that yields a decrease of the skin friction instead of the typical increase. Hence, drag changes due to riblets can be characterized by a quantity often employed to describe the effects of wall roughness: the vertical shift induced in the logarithmic portion of the mean velocity profile when plotted in the law-of-the-wall form. This shift, positive (upward) in case of drag reduction and physically interpreted as a consequence of the thickening of the viscous sublayer \citep{choi-1989}, implicitly contains the $Re$-dependency of $C_f$ through the friction velocity used for the adimensionalization. Thus, if $Re$ is large enough for the logarithmic law to be valid in a $Re$-independent form, the amount of this shift is constant with $Re$, and is the preferred way to characterize the ability of riblets to reduce drag, as clearly advocated by \cite{spalart-mclean-2011}.

Turning now our attention to active, wall-based techniques for skin-friction drag reduction, a lively debate is taking place in the scientific community regarding the high-$Re$ behaviour of the last generation of techniques, in particular the open-loop ones that promise large benefits with the advantage of reasonable implementation complexity. Such techniques operate by enforcing suitable temporal and spatial distributions of velocity perturbations at the wall, and have been shown to be able to relaminarize an otherwise turbulent flow, with an energy cost that can be significantly smaller than the energy savings. Assessing their potential for achieving sizeable benefits at high $Re$ is obviously important to motivate further research in this field.

The specific case of spanwise-forcing techniques for drag reduction includes the well-known spanwise-oscillating wall \citep{jung-mangiavacchi-akhavan-1992} and its generalizations made by the spanwise-travelling waves \citep{du-symeonidis-karniadakis-2002} and streamwise-travelling waves \citep{quadrio-ricco-viotti-2009}; although most of the available information concerns internal flows, numerical experiments \citep{skote-2011, lardeau-leschziner-2013, skote-2013, mishra-skote-2015} have shown that the forcing is effective in external flows too. The streamwise-travelling waves, in particular, yield larger maximum drag reduction and, more importantly, improved energetic efficiency, which is essential if drag reduction is motivated by the need to save energy.
Laboratory implementations of such techniques range from proof-of-principle experiments \citep{auteri-etal-2010} to experiments with novel actuation technologies, such as electroactive polymers \citep{gouder-potter-morrison-2013} or plasma actuators, specifically addressing the problem of actuation efficiency \citep{gatti-etal-2015}.
Thanks to the wealth of data available for this class of forcing, it is known that the impressive low-$Re$ performance, i.e. 58\% drag reduction and 28\% net energy saving at $Re_\tau=200$ in a turbulent channel flow \citep{quadrio-ricco-viotti-2009}, and flow relaminarization at $Re_\tau=200$ in a turbulent pipe flow \citep{xie-2014}, does indeed degrade with $Re$. 

The drag reduction rate $\R$ is often considered to decrease with the Reynolds number following a power law, i.e. $\R \sim Re_\tau^{-\gamma}$ with the exponent $\gamma$ determined empirically.  Early $Re$-dependency studies were typically based upon a large parameter study carried out at low $Re$ and given forcing amplitude, where the best combination of parameters for drag reduction was first identified; this sole or a few sets of parameters were then tested at higher $Re$, and the performance drop measured under the assumption that the control parameters identifying the optimal point scale in viscous wall units. However, as discussed by \cite{quadrio-2011}, the assumption of viscous scaling not only needs to be verified, but also implies a choice in selecting the velocity scale: in drag reduction studies, besides the friction velocity $u_{\tau,0}$ of the non-controlled flow, an additional friction velocity is available to build viscous units, namely the actual friction velocity $u_\tau$ of the drag-reduced flow. In this paper, we will indicate with the customary $+$ superscript the ``reference'' wall units, whereas a $*$ superscript will indicate the "actual" wall units. Limiting to the literature where more than one value of $Re$ was considered, quantities have been scaled through either reference $+$ units \citep{quadrio-ricco-viotti-2009, touber-leschziner-2012,gatti-quadrio-2013,hurst-yang-chung-2014} or actual actual $*$ units \citep{moarref-jovanovic-2012}. The choice of nondimensionalization is particularly delicate and may cause spurious $Re$ effects, since $\R$ is function of all control parameters including $Re$.

A recent study that thoroughly describes the literature relevant to the higher-$Re$ behaviour of spanwise-forcing techniques is that by \cite{hurst-yang-chung-2014}. \cite{choi-xu-sung-2002} considered the oscillating wall for $Re_\tau=100, 200$ and $400$ and, although not addressing the Reynolds dependency directly, provided results implying $\gamma=0.2-0.4$. \cite{ricco-quadrio-2008} considered the oscillating wall at $Re_\tau=200$ and $Re_\tau=400$ and measured $\gamma=0.21$ for the point with largest drag reduction, although the rate of drecrease appeared to depend on the parameters of the oscillating wall. \cite{quadrio-ricco-viotti-2009} for the streamwise-travelling waves measured $\gamma=0.24$. \cite{touber-leschziner-2012} went up to $Re_\tau=1000$ for the oscillating wall and suggested $\gamma=0.20$. This is a rather strong decrease rate: with $\gamma=0.24$, the 58\% drag reduction of the travelling waves at $Re_\tau=200$ would become a mere 7\% at $Re_\tau=10^6$. Note that these figures do not account yet for the energy cost of the active control. 

However, there are indications that the picture might be more complicated and possibly not as negative. First of all, the few experimental data available \citep{choi-graham-1998, ricco-wu-2004} observe small, if any, $Re$ effect. Moreover, numerical studies based on alternative approaches suggest a different scenario. \cite{duque-etal-2012} solved the linearized Naver--Stokes equations to study how the growth of the near-wall streaks is affected by the travelling waves; they found no significant $Re$-effect when comparing the growth between $Re_\tau=200$ and $Re_\tau=2600$. They also observed that, at least at $Re_\tau=200$ where DNS data for the drag change are available, streak amplification correlates well with drag reduction. Thus, the implicit message is that drag reduction might be mildly affected by increasing $Re$, similarly to streak growth. \cite{moarref-jovanovic-2012} developed a model-based approach for studying spanwise-wall oscillations. They used eddy-viscosity-enhanced linearization of the turbulent flow with control in conjunction with turbulence modelling to determine skin-friction drag without resorting to heavy simulations. Unfortunately, their method relies upon the availability of statistical results from a DNS of the uncontrolled flow at the same $Re$, and does not lend itself to being used at very high Reynolds numbers. However, up to $Re_\tau=1000$ they found $\gamma=0.15$ for the oscillating wall. \cite{belan-quadrio-2013} studied turbulent drag reduction in conjunction with the RANS equations, and carried out an asymptotic analysis to extrapolate the performance of streamwise-travelling waves at high $Re$. Their finding is that at $Re_\tau=$20,600 drag reduction decreases only by 15\% from the value at $Re_\tau=200$, which is equivalent to $\gamma=0.04$.

The recent study by \cite{gatti-quadrio-2013} began to shed some light upon these contrasting indications. Their numerical study, based on DNS applied to rather small computational domains, went up to $Re_\tau=1000$ with a few points at $Re_\tau=2000$, and confirmed that maximum drag reduction at the considered forcing amplitude indeed decreases quickly with $\gamma \approx 0.2$. However, \cite{gatti-quadrio-2013} were able to carry out a larger parameter study at $Re_\tau=1000$ too, and evaluated $\gamma$ at several points in the space of control parameters, finding that $\gamma$ is a strong function of the parameters which define the forcing, and that the performance drop is more pronounced where drag reduction at low $Re$ is maximum. The results obtained by \cite{gatti-quadrio-2013} were confirmed by \cite{hurst-yang-chung-2014} with a less detailed parameter study but with the added reliability of using DNS with full-size computational domains. Data at $Re_\tau=200, 400, 800$ and $1600$ confirmed that $\gamma$ depends on the flow control parameters, typically ranging between 0.1 and 0.5. 

Overall, this calls for a deeper understanding and generalization of these findings, as the outlook for practical applications would be strongly affected. Hence, the goal of the present paper is to further address the high-$Re$ behavior of spanwise forcing techniques. Two DNS databases are built to characterize the streamwise-travelling waves in terms of drag reduction and net energy savings at two, well separated values of the Reynolds number: $Re_\tau=200$ and $Re_\tau=1000$. The parameter study is extremely large and is made by 4020 DNS calculations on small domain sizes, plus 20 additional cases with a larger domain size. For the first time the amplitude of the travelling waves is systematically studied together with their wavelength and speed: this will for the first time allow for a clear interpretation of the results in terms of the actual forcing intensity. Details of the numerical procedures and the computational parameters are given in \S\ref{sec:method}, with particular attention to the spatial discretization issue: the employed computational domain, like in \cite{gatti-quadrio-2013}, is smaller than the usual to reduce the computational cost, and monitoring the uncertainty of the results is essential. The results are then presented and discussed in \S\ref{sec:results}, while in \S\ref{sec:re-effect} the drag reduction data are examined in terms of $\gamma$, to show that this quantity is not particularly well suited to describe the $Re$-effect. The vertical shift in the logarithmic portion of the mean velocity profile is then shown to be a more robust alternative. An analytical relation is eventually developed to predict drag reduction at arbitrarily high values of $Re$ based on low-$Re$ information.

\section{Method}
\label{sec:method}

\begin{figure}
\centering
\includegraphics{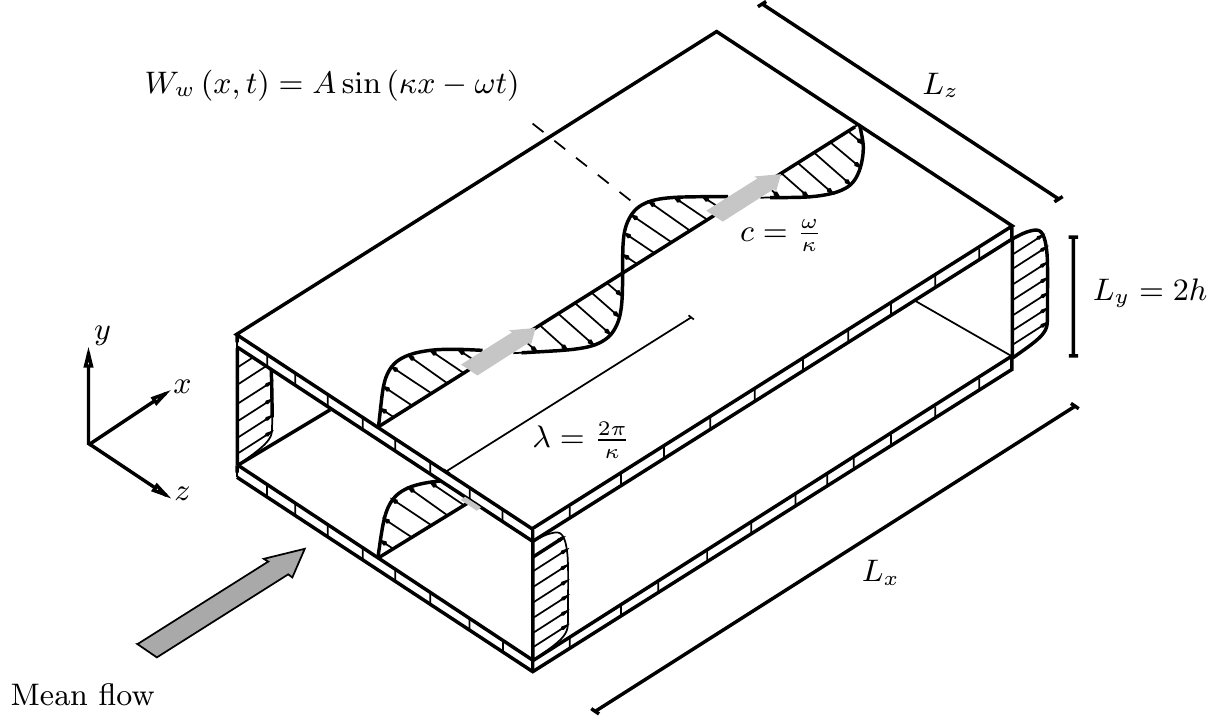}
\caption{Schematic of a turbulent channel flow modified by streamwise-travelling waves of spanwise wall velocity, with amplitude $A$, streamwise wavenumber $\kappa$ and angular frequency $\omega$. $\lambda$ is the streamwise wavelength and $c$ is the phase speed of the waves. $L_x$, $L_y=2h$ and $L_z$ are the dimensions of the computational domain in the streamwise, wall-normal and spanwise directions, respectively.}
\label{fig:traveling-sketch}
\end{figure}

Direct Numerical Simulations (DNS) of the turbulent flow in a doubly periodic channel are used to study the effect of the Reynolds number $Re$ on the reduction of turbulent drag achieved by streamwise-travelling waves of spanwise wall velocity \citep{quadrio-ricco-viotti-2009}, imposed on both walls as the boundary condition:
\begin{equation}
W_w(x,t) = A \sin (\kappa x - \omega t) .
\label{eq:tw}
\end{equation}

In the above expression for the wall forcing, $W_w$ is the spanwise velocity enforced at the wall, $A$ is the amplitude of the forcing, $\kappa$ is the streamwise wavenumber and $\omega$ is the angular frequency. $x$ and $t$ are the streamwise coordinate and time respectively. The forcing, sketched in figure \ref{fig:traveling-sketch}, consists in a wall distribution of streamwise-modulated waves of the spanwise ($z$) velocity component with wavelength $\lambda = 2 \pi / \kappa$ and period $T= 2 \pi / \omega$, which travel at speed $c = \omega / \kappa$ forward ($c>0$) or backward ($c<0$) with respect to the direction $x$ of the mean flow. The three independent parameters (for example $A$, $\kappa$, $\omega$) of the control law (\ref{eq:tw}) combined with the Reynolds number $Re$ define a 4-dimensional parameter space, whose complete investigation represents a computational challenge. 

The simulations have been run with the solver for the incompressible Navier--Stokes equations developed by \cite{luchini-quadrio-2006}, and adapted in this work to run on a Blue Gene/Q system at the CINECA computing centre, where most of the calculations were carried out. Simulations enforcing either a constant flow rate (CFR) or a constant pressure gradient (CPG) \citep[see][for details]{hasegawa-quadrio-frohnapfel-2014, quadrio-frohnapfel-hasegawa-2016} are considered. The aim is to obtain and compare two comprehensive sets of cases at $Re_\tau=200$ and $Re_\tau=1000$, where $Re_\tau = u_\tau h / \nu$ is the Reynolds number based on the channel half-height $h$, the friction velocity $u_\tau$ of the uncontrolled flow and the kinematic viscosity $\nu$ of the fluid. This is enforced directly in the CPG cases where the value of $Re_\tau$ is specified as an input parameter, while in the CFR cases the imposed flow rate (hence the input value of $Re_b=U_b \,2h / \nu$) leads to values of $Re_\tau$ which only approximately correspond to the target. In the following, for ease of discussion we will conventionally refer to the two casesets as the low-$Re$ (or $Re_\tau=200$) caseset and the high-$Re$ (or $Re_\tau=1000$) caseset.

In both casesets, the initial condition is that of an uncontrolled turbulent flow, and care is taken to begin the statistical analysis after properly discarding the initial transient (of duration up to 1,000 viscous time units) where the control leads the flow towards a reduced level of friction drag. The spatial resolution in wall units is always better than $\Delta x^+ = 12.3$ and $\Delta z^+=6.1$ (or $\Delta x^+=8.2$ and $\Delta z^+ = 4.1$ if the additional modes used to completely remove the aliasing error are considered). $\Delta y^+$ smoothly varies from $\Delta y^+ \approx 1$ near the wall to $\Delta y^+ \approx 7$ at the centerline. In the CFR cases, the spatial resolution in every direction strongly improves with drag reduction and the related drop in $Re_\tau$. Time integration is carried out with a partially implicit approach, with a Crank-Nicolson scheme for the viscous terms and a third-order Runge--Kutta scheme for the convective terms. The CFL number is set at unity, well below the stability limit of the temporal integration scheme; the consequent average size of the timestep is always below $\Delta t^+ = 0.17$ for the low-$Re$ cases, and below $\Delta t^+ = 0.1$ for the high-$Re$ cases. The duration of the simulation obviously affects the quality of the estimate of the mean value of drag: the total integration time is thus adjusted in order to reach an acceptably small uncertainty over the whole dataset. The integration time is at least 24,000 viscous time units, and in certain cases it increases up to 80,000 viscous units. For each value of $Re$, the computational study considers two distinct sets of simulations, described below, details of which are reported in table \ref{tab:discretization-parameters}.

\begin{table}
\begin{tabular*}{1.0\textwidth}{@{\extracolsep{\fill}} l r r r r r r r c}
Type & $N_{\mathrm{cases}}$ & $Re_\tau$ & $Re_b$ & $L_x/h$& $L_z/h$ & $L_x^+$& $L_z^+$& $N_x \times N_y \times N_z$\\[10pt]

CFR & 1530 &  199.0 &  6347  & 1.59$\pi$ & 0.80$\pi$ &   995   &  498   &  $96 \times 100 \times  96$  \\	
CFR &  480 &  199.2 &  6347  & 2.05$\pi$ & 1.02$\pi$ &  1281   &  640   & $128 \times 100 \times 128$  \\
CFR & 1530 &  905.6 & 39333  & 0.32$\pi$ & 0.16$\pi$ &   906   &  453   &  $96 \times 500 \times  96$   \\
CFR &  480 &  948.3 & 39333  & 0.43$\pi$ & 0.22$\pi$ &  1290   &  645   & $128 \times 500 \times 128$  \\
\hline
CFR &    5 &  199.9 &  6361  & $4\pi$    & $2\pi$    &  2512   & 1256   & $ 256 \times 128 \times  256$  \\	
CPG &    5 &  200.0 &  6358  & $4\pi$    & $2\pi$    &  2513   & 1257   & $ 256 \times 128 \times  256$  \\
CFR &    5 &  998.7 & 39990  & $4\pi$    & $2\pi$    & 12550   & 6275   & $1024\times  500 \times 1024$ \\
CPG &    5 & 1000.0 & 39992  & $4\pi$    & $2\pi$    & 12566   & 6283   & $1024\times  500 \times 1024$ \\	
\end{tabular*}
\caption{Details of the small-box (upper half) and large-box (lower half) simulations. Every caseset is detailed in terms of simulation type (CFR or CPG), number of cases $N_{\mathrm{cases}}$, values of bulk Reynolds number $Re_b$ and friction Reynolds number $Re_\tau$, length and width of the computational domain in inner and outer units, number of Fourier modes in the homogeneous directions (additional modes are used for dealiasing, according to the 3/2 rule) and collocation points in the wall-normal direction.}
\label{tab:discretization-parameters}
\end{table}

\subsection{The small-box large database}
\label{sec:smallbox}
The first set (upper half of table \ref{tab:discretization-parameters}) is a parameter study designed to produce a massive database of drag reduction data (4020 cases overall); the parameter space includes the forcing wavenumber $\kappa$, the forcing angular frequency $\omega$ and, for the first time, the forcing amplitude $A$ too. For each $Re$, six values of $A^+=\left\{ 2,\, 4.5,\, 5.5,\, 7,\, 12,\, 20\, \right\}$, a wide frequency range $-0.5 \leq \omega^+ \leq 1$, and 13 different values of $\kappa^+$ are considered, between the limiting case $\kappa^+=0$ when the control law (\ref{eq:tw}) reduces to the classic spanwise-oscillating wall and the maximum considered wavenumber $\kappa^+ = 0.05$. Previous knowledge of the drag reduction pattern is exploited to focus on the interesting regions of the parameter space through a non-uniform distribution of the simulation points. Values of $A^+$ are chosen to describe reasonably well the region for maximum energetic efficiency near $A^+=5$, to include the point $A^+=12$ that corresponds to most available information \citep{quadrio-ricco-viotti-2009, touber-leschziner-2012, hurst-yang-chung-2014}, and to include an additional point at larger $A^+$ to better characterize the absolute maximum of drag reduction. Availability of data at different $A^+$ is essential for appreciating the difference, discussed below, between ``reference $u_\tau$'' or $+$ scaling and ``actual $u_\tau$'' or $*$ scaling.

For this set of calculations, carried out under the CFR condition, a relatively small computational domain (whose dimensions are kept constant in wall units when $Re$ is increased) is employed: the consequent savings in computing time are key to make this huge parameter study possible. For each value of $Re_b$, adjusted to yield the target value of $Re_\tau$ as nearly as practical, two subsets of simulations are reported in table \ref{tab:discretization-parameters}, which differ for the slightly different value of the length $L_x$ of the computational domain. Since the wave length of the wall forcing is constrained to integer submultiples of the box length, this mitigates the quantization effect and allows a better investigation of the lower forcing wavenumbers. In changing box size, the aspect ratio of the computational domain is always kept constant at $L_x = 2 L_z$, and the number of Fourier modes in the wall-parallel directions is adjusted as to keep the spatial resolution in wall units unchanged. 

\subsection{The large-box small database}
\label{sec:largebox}
The second set of simulations (lower half of table \ref{tab:discretization-parameters}) employs a larger domain size, whose dimensions are kept constant in outer units when $Re$ is increased. Owing to the larger computational cost, only a few representative cases are computed. For both $Re$ we consider the reference uncontrolled case, and four other cases at the amplitude $A^+=7$, which is a control amplitude near the largest net savings. One case is for the oscillating wall at nearly optimal period $T^+=75$, one case with oscillating wall at the larger period $T^+=250$, one case with travelling waves with large drag reduction ($\omega^+=0.0239$ and $\kappa^+=0.01$) and one case for travelling waves with drag increase ($\omega^+=0.12$ and $\kappa^+=0.01$). Each case is run under both CFR and CPG (and for the latter the forcing parameters listed above are to be intended in actual wall units), for a total of 20 simulations featuring the larger computational domain.

\subsection{Domain size and uncertainty}
\label{sec:uncert}
For the simulations described in \S\ref{sec:smallbox}, the computational domain is rather small, with a streamwise length of the order of 1000 wall units (one half of that for the domain width). Such domains, which imply a reduced computational cost per timestep, have been already used in the past for similar studies. Although several times larger than the minimal domain shown by \cite{jimenez-moin-1991} to be capable of sustaining the near-wall turbulent cycle, the present computational domain is not large enough to guarantee flow statistics truly independent from spatial truncation. The large outer structures that become more and more important as $Re$ increases do not entirely fit into such a small domain, the mean velocity in the outer region progressively deviates from the correct profile, and even the prediction of wall friction is affected by an error. Trace of this can be observed in table \ref{tab:discretization-parameters}, where at high $Re$ a 35\% increase in box length brings about a 5\% change in $Re_\tau$.

However, \cite{gatti-quadrio-2013} demonstrated that reliable information concerning drag reduction can still be extracted from such simulations. In fact, most of the inaccuracy in predicting friction cancels out when the two friction coefficients (with and without control) are subtracted to compute drag reduction. It must be noted that the much shorter computing time per timestep comes at the cost of a larger number of timesteps required to obtain meaningful temporal averages of spatially-averaged quantities, which present larger temporal fluctuations as a consequence of the lesser contribution of the spatial average \citep{jimenez-moin-1991}. \cite{lozanoduran-jimenez-2014} have shown that the standard deviation of the spatially-averaged skin friction is roughly inversely proportional to the square root of the area of the computational box. Hence, it is imperative to monitor the statistical uncertainty of the main quantities, and to increase the averaging time (i.e. the duration of the simulations) until the uncertainty reduces to an acceptable level. 

In the present paper, the main quantity of interest is the drag reduction rate $\R$, as defined in Eq.~(\ref{eq:DRR}). Its uncertainty $\delta \R$ is written by propagating the standard uncertainty of the skin-friction coefficients for the uncontrolled and controlled flows, and by assuming they are independent variables:
\begin{equation}
\delta \R = \frac{C_f}{C_{f,0}} 
\sqrt{ \left( \frac{\delta C_f}{C_f} \right)^2 
       + \left( \frac{\delta C_{f,0}}{C_{f,0}} \right)^2  }
\label{eq:uncertainty}
\end{equation}
and then extended to a 95\% confidence interval. The standard uncertainties $\delta C_{f,0}$ and $\delta C_f$ are evaluated with the procedure described by \cite{oliver-etal-2014}, by calculating an integral timescale via an autoregressive method. This is slightly different from the strategy employed by \cite{gatti-quadrio-2013}, who estimated the timescale differently, and it is interesting to note that the two methods end up with essentially the same result. For instance, when applied to the time history of skin-friction in an unmanipulated channel flow at $Re_\tau=200$, with domain size $L_x = 4 \pi h, L_z = 2 \pi h$ and an integration time of 4480 $h / U_b$, the present method yields a relative standard uncertainty of 0.70\% while the method by \cite{gatti-quadrio-2013} yields the slightly lower value of 0.66\%. Similar values of 0.61\% and 0.64\% are obtained respectively with the well-known method of non-overlapping batch means \citep{schmeiser-1982}, which is related to the procedure followed by \cite{hoyas-jimenez-2008}, and the similar method proposed by \cite{mockett-knacke-thiele-2010}, which relies on the known statistical properties of white noise. The present method provides the most conservative estimate. 

The uncertainty on the value of other quantities, such as the mean velocity profile or the control performance estimators presented below, is computed analogously. Hereinafter the uncertainty is always reported in the text when applicable and represented with error bars or shading in all the relevant figures. 

\section{Results}
\label{sec:results}

\subsection{Validation for the oscillating wall}

\begin{figure}
\centering
\includegraphics{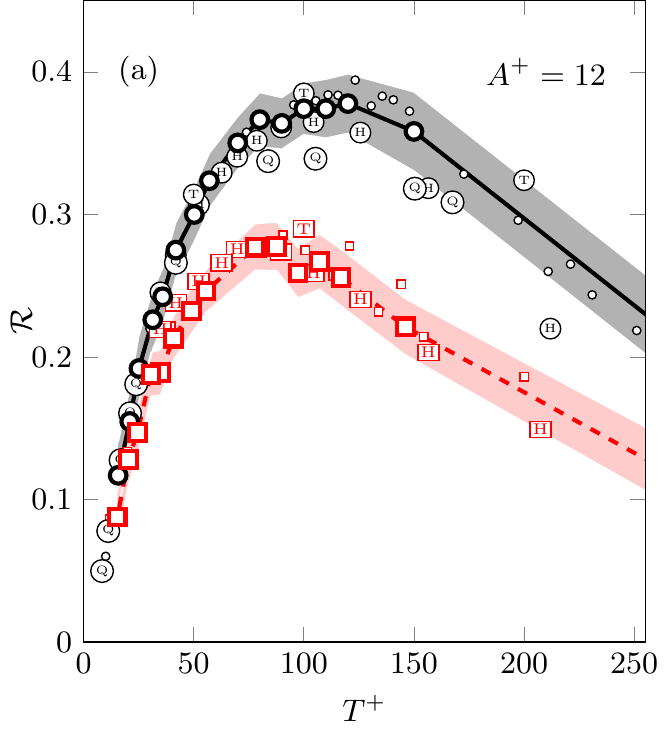}%
\includegraphics{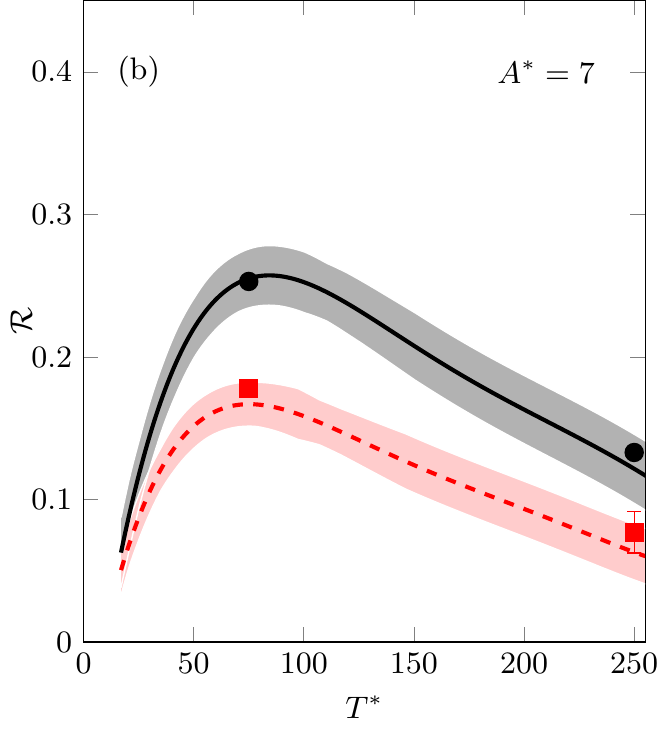}
\caption{Drag reduction rate $\R$ versus period of wall oscillation $T^+$ at $A^+=12$ in reference scaling (a) and versus $T^*$ at $A^*=7$ in actual scaling (b). Black (darker) color identifies low-$Re$ data, and red (lighter) color corresponds to high-$Re$ data. Present data are identified by large symbols connected with thick lines, with the shaded area representing the corresponding interval at 95\% confidence level. In panel (a), small open symbols are from \cite{gatti-quadrio-2013}, while symbols with letters identify literature data: Q, \cite{quadrio-ricco-2004}; T, \cite{touber-leschziner-2012}; H, \cite{hurst-yang-chung-2014} with the higher Reynolds data at $Re_\tau=800$. In panel (b), present data are represented by thick lines only, because these are obtained through linear interpolation, while symbols are present data for large-box simulations.}
\label{fig:OWcomp}
\end{figure}

We show first the results for the temporal oscillation of the wall (i.e. $\kappa=0$), that can be used as a further validation of the present dataset, thanks to the availability of several literature data at various $Re$. Figure \ref{fig:OWcomp} shows in the left panel the drag reduction rate against the period of wall oscillation $T^+$ for oscillations with an amplitude of $A^+=12$ in reference inner scaling. At $Re_\tau=200$ the present results agree very well with those from \cite{gatti-quadrio-2013}, who employed a comparable domain size, as well as with the full-size simulations of \cite{touber-leschziner-2012} and \cite{hurst-yang-chung-2014}. An optimal oscillation period exists at $T^+=100-125$ \citep{baron-quadrio-1996, yakeno-hasegawa-kasagi-2014}, which maximizes at every $A^+$ the interaction of the oscillating Stokes layer of proper thickness with the convecting turbulence structures. At optimal values of $T^+$ and $A^+=12$, for instance, \cite{touber-leschziner-2012} observed $\R=0.385$, \cite{hurst-yang-chung-2014} $\R=0.364$ and \cite{gatti-quadrio-2013} $\R=0.384$. All agree well with the present value of $0.374 \pm 0.018$. The only exception is the data point at $T^+\approx 200$ by \cite{hurst-yang-chung-2014}, which lies well below the other datasets. (It is possible that their averaging time was chosen too small and/or not as an integer multiple of the oscillation period: at large $T^+$ the wall friction exhibits significant periodic oscillations. Indeed, travelling waves data by \cite{hurst-yang-chung-2014}, which are not affected by this problem, agree very well with present and other results: see figure \ref{fig:Rmax} below.) The comparison against the reference dataset by \cite{quadrio-ricco-2004} confirms the slight (less than 0.02) overestimation of $\R$ in a narrow region close to the optimal period $T^+ \approx 100$, already known \citep[cfr. the discussion of figures 5 and 6 in][]{gatti-quadrio-2013} and attributed to domain size effects; this is limited to the region of largest drag reduction ($\R>0.3$) and becomes milder for streamwise-travelling waves ($\kappa > 0$). The agreement between the present dataset and the available literature data is excellent also at $Re_\tau=1000$. The datapoints by \cite{hurst-yang-chung-2014} lie slightly above the present results for small oscillation periods, but it must be recalled that they are computed at the lower $Re_\tau=800$. As already known \citep{gatti-quadrio-2013, hurst-yang-chung-2014}, the optimal oscillation period decreases with $Re$ and becomes $T^+\approx 75$ at $Re_\tau=1000$. At this value of $T^+$ a drag reduction of $0.277 \pm 0.016$ is measured, which is very close to $\R=0.275$ measured by \cite{hurst-yang-chung-2014} and $\R=0.29$ reported by \cite{touber-leschziner-2012} at $T^+=100$. 

Figure \ref{fig:OWcomp}(b) shows $\R$ versus $T^*$ at constant $A^* = 7$, i.e. with actual viscous scaling, which would require data from simulations driven at CPG. The present small-box simuations are indeed computed at CFR, but the availability of data at several values of $A^+$ makes a $*$ scaling possible through interpolation. As far as we know, no such data exist for $Re_\tau=1000$ or nearby; the only data we can compare with are the large-domain CPG  simulations described in table \ref{tab:discretization-parameters}, and this motivates the choice of the amplitude $A^*=7$. At $Re_\tau=200$ the optimal period is $T^* \approx 75$, where a drag reduction of $0.255 \pm 0.020$ is achieved. This value is very close to $0.253  \pm 0.002$ of the large-domain data point at the same $T^*$. Good agreement between the present full-size DNS and reduced domain database is evident also at $Re_\tau=1000$, where at $T^*=75$ the former achieved $0.167 \pm 0.015$ and the latter $0.175 \pm 0.005$. The shift of the optimal $T^*$ with $Re$ is, if present at all, much milder than that of $T^+$, suggesting that actual inner scaling might be more appropriate. A possible rationale to explain the existence of the optimal oscillation period can be found in \cite{yakeno-hasegawa-kasagi-2014}.

\subsection{The whole dataset at a glance}

\begin{figure}
\centering
\includegraphics{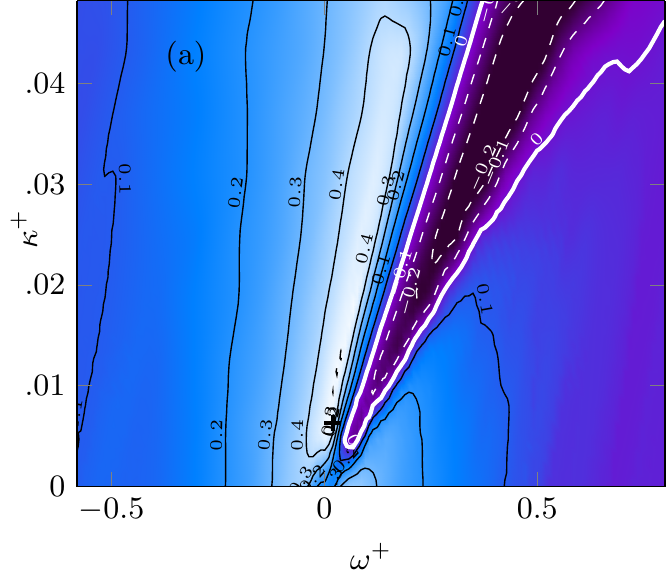}\includegraphics{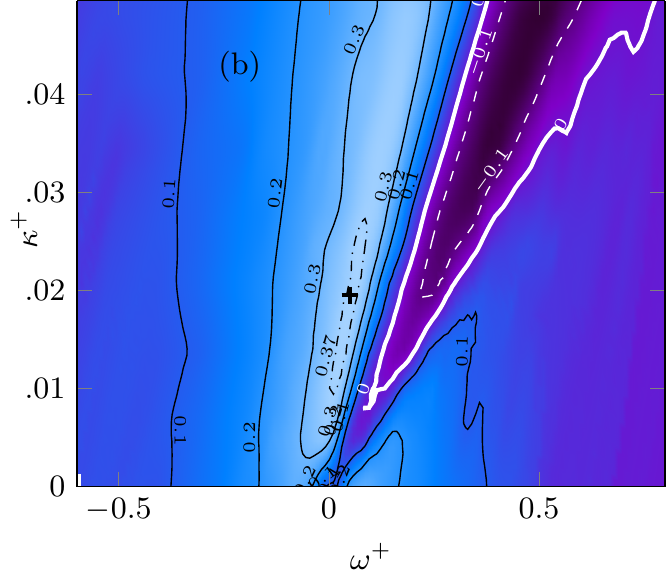}
\caption{Map of drag reduction $\R_{A^+}$ at constant forcing amplitude for streamwise travelling waves at $A^+=12$, for $Re_\tau=200$ (a) and $Re_\tau=1000$ (b). Contours are spaced by 0.1, negative contours are dashed. The thick white line corresponds to zero drag change while the dash-dotted line is the locus of points where $\R = \R_{m,A^+}-\delta \R$, and the cross indicates the position of $\R_{m,A^+}$.}
\label{fig:R-A12}
\end{figure}

Travelling-wave results are now presented as in \cite{quadrio-ricco-viotti-2009} and \cite{hurst-yang-chung-2014}, by plotting drag reduction maps in the space of the control parameters $\left(\kappa^+, \omega^+ \right)$ at $A^+=12$. The data just discussed in figure \ref{fig:OWcomp} are those lying on the $\kappa=0$ horizontal axis. Note that to produce such maps, first the raw results are interpolated on a finer Cartesian grid, without any smoothing or outliers removal, and then contour lines are drawn on the Cartesian grid. Using * scaling also implies interpolation along the third axis (forcing amplitude). Figure \ref{fig:R-A12} compares the drag reduction rate $\R$ at $Re_\tau=200$ and $Re_\tau=1000$. The map at $Re_\tau=200$ reproduces the well known shape described by \cite{quadrio-ricco-viotti-2009}, with a maximum $\R$ at given $A^+=12$ of $\R_{m,A^+}=0.500 \pm 0.015$ located at $\left( \omega^+, \kappa^+ \right)=\left(0.0195, 0.0063 \right)$, which agrees with $\R_{m,A^+}=0.48$ at $\left(0.018, 0.005 \right)$  measured by \cite{quadrio-ricco-viotti-2009}  and $\R_{m,A^+}=0.5$ at $\left(0.025, 0.008 \right)$ obtained by \cite{hurst-yang-chung-2014}. The triangular drag-increasing region associated \citep{quadrio-ricco-viotti-2009} to a lock-in phenomenon between the travelling waves and the convecting turbulence structures is clearly evident and well captured. The side-by-side comparison with data at $Re_\tau=1000$ clearly shows that, consistent with all previous information, the forcing becomes less effective as $Re$ grows, since both the largest drag reduction and drag increase weaken. $\R_{m,A^+}$ decreases to $0.388 \pm 0.014$ and its location in the plane moves along the drag reduction ridge towards higher frequencies and wavenumbers to $\left( \omega^+, \kappa^+ \right)=\left(0.05, 0.0195 \right)$. It must be observed, however, that the precise location of $\R_{m,A^+}$ in the $\left( \omega^+, \kappa^+ \right)$ plane is not a robust information, since the region of maximum drag reduction presents a rather flat peak. This is particularly true at $Re_\tau=1000$, where the contour level drawn for $\R = 0.388 - 0.014$, i.e. $\R_{m,A^+}$ minus its uncertainty, is an elongated region which almost encloses the low-$Re$ maximum. However, the decrease of $\R_{m,A^+}$ at larger $Re$ as well as its shift towards higher frequencies and wavenumbers is clear, and it agrees with previous evidence by \cite{gatti-quadrio-2013} and by \cite{hurst-yang-chung-2014}, with the latter being based on fewer data at $Re_\tau=800$ but computed on a larger computational domain.





\begin{figure}
\centering
\includegraphics{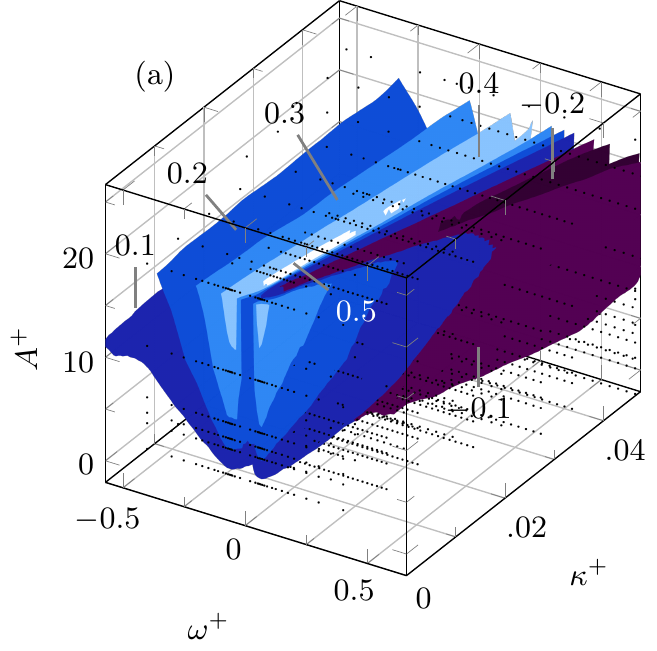}\includegraphics{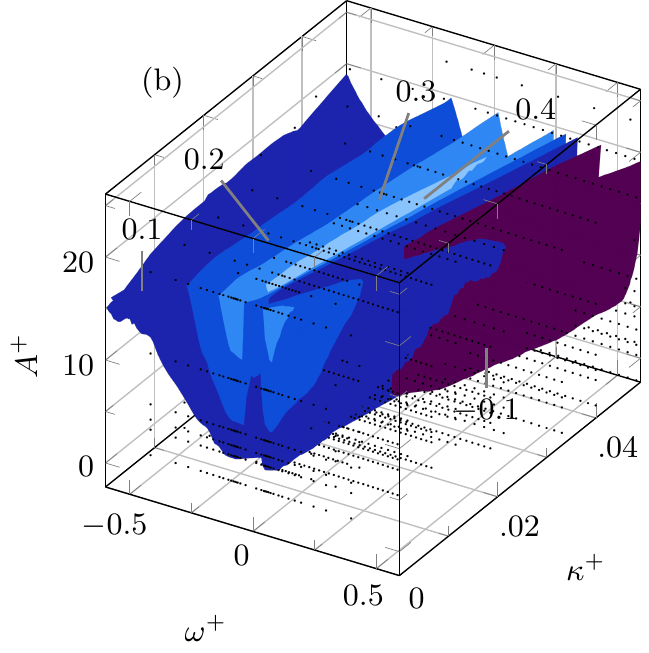}
\caption{Isosurfaces of drag reduction $\R$ in the three-dimensional parameter space $\left(\omega^+, \kappa^+, A^+ \right)$ for $Re_\tau=200$ (a) and $Re_\tau=1000$ (b). Isosurface from dark to light range from $\R=-0.2$ to 0.5 in steps of 0.1. The cloud of dots represents the 2010 data points where, at each $Re$, a DNS has been carried out. 
\label{fig:3DR}}
\end{figure}

\begin{figure}
\centering
\includegraphics{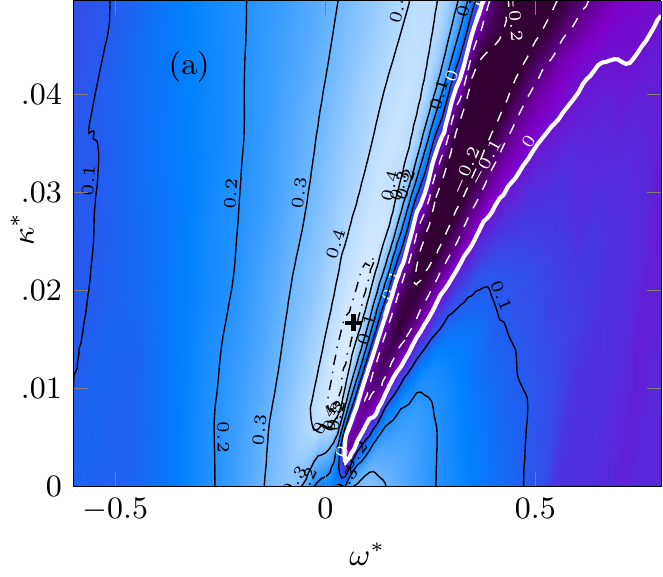}\includegraphics{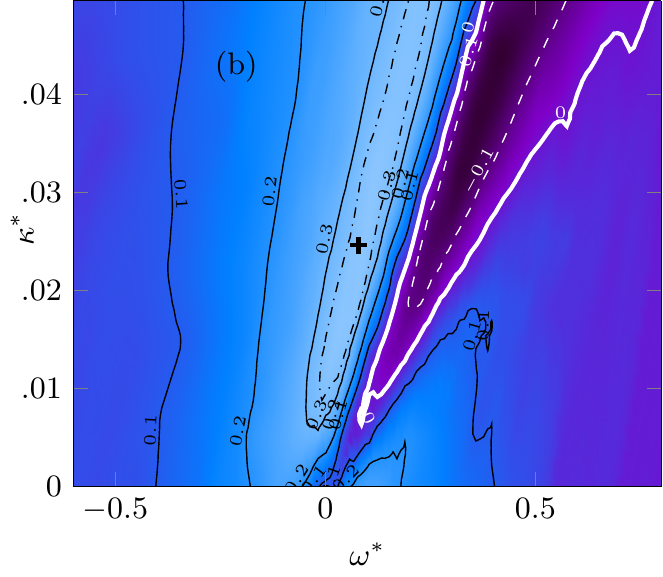}
\caption{Map of drag reduction $\R_{A^*}$ at constant forcing amplitude for streamwise travelling waves at $A^*=12$, for $Re_\tau=200$ (a) and $Re_\tau=1000$ (b). Lines and symbols as in figure \ref{fig:R-A12}.}
\label{fig:r-A12}
\end{figure}

The present database contains much more information than that shown in figure \ref{fig:R-A12}, because several values of forcing amplitude are considered. Figure \ref{fig:3DR} shows the entire dataset at a glance in the three-dimensional parameter space  $\left( \omega^+, \kappa^+, A^+ \right)$, in comparative form between the two Reynolds numbers. The three-dimensional field of $\R$ is visualised via isosurfaces of $\R = const$, and the clouds of tiny dots in the figure indicates the 4020 points where a DNS has been performed. Such an extensive database makes the description of $\R$ with actual inner scaling possible. In the past, only few channel flow studies \citep[for example][]{ricco-etal-2012} were performed at CPG, thus ensuring that drag reduction does not lead to changes in $Re_\tau$. When CFR is employed, a significant drop in the friction velocity and $Re_\tau$ due to drag reduction implies that viscous units change, and in particular that the forcing for a given $A^+$ becomes stronger in physical units than the forcing for the same given value of $A^*$. Using + scaling is thus equivalent to employing outer scaling. Hence, the choice between $+$ and $*$ scaling leads to different scenarios \citep{quadrio-2011}. Since in the drag reduced state only the actual viscous units are relevant, $*$ scaling should be adopted when comparing results obtained at different $Re$. The present data have been computed on $A^+=const$ planes, but they can be easily interpolated to produce figure \ref{fig:r-A12}, where a map of $\R$ in the $\left( \omega^*, \kappa^* \right)$ plane at $A^*=12$ is shown. At $Re_\tau=200$, the maximum drag reduction at constant $A^*$ is $\R_{m,A^*}=0.453 \pm 0.015$ at $\left( \omega^*, \kappa^* \right)=\left(0.0167, 0.067 \right)$, to be compared with $\R_{m,A^+}=0.500 \pm 0.015$ at $\left( \omega^+, \kappa^+ \right)=\left(0.0195, 0.0063 \right)$. This agrees with data at same $A^*$ and $Re$ by \cite{quadrio-ricco-2011}, who drove the channel at CPG and found $\R_{m,A^*}=0.45$ at $\left( \omega^*, \kappa^* \right)=\left(0.012, 0.045 \right)$. At the higher $Re_\tau=1000$, $\R_{m,A^*}=0.343 \pm 0.019$ at $\left( \omega^*, \kappa^* \right)=\left(0.0246, 0.0789 \right)$ is observed instead of $\R_{m,A^+} = 0.388 \pm 0.014$ at $\left( \omega^+, \kappa^+ \right)=\left(0.05, 0.0195 \right)$.

\begin{figure}
\centering
\includegraphics{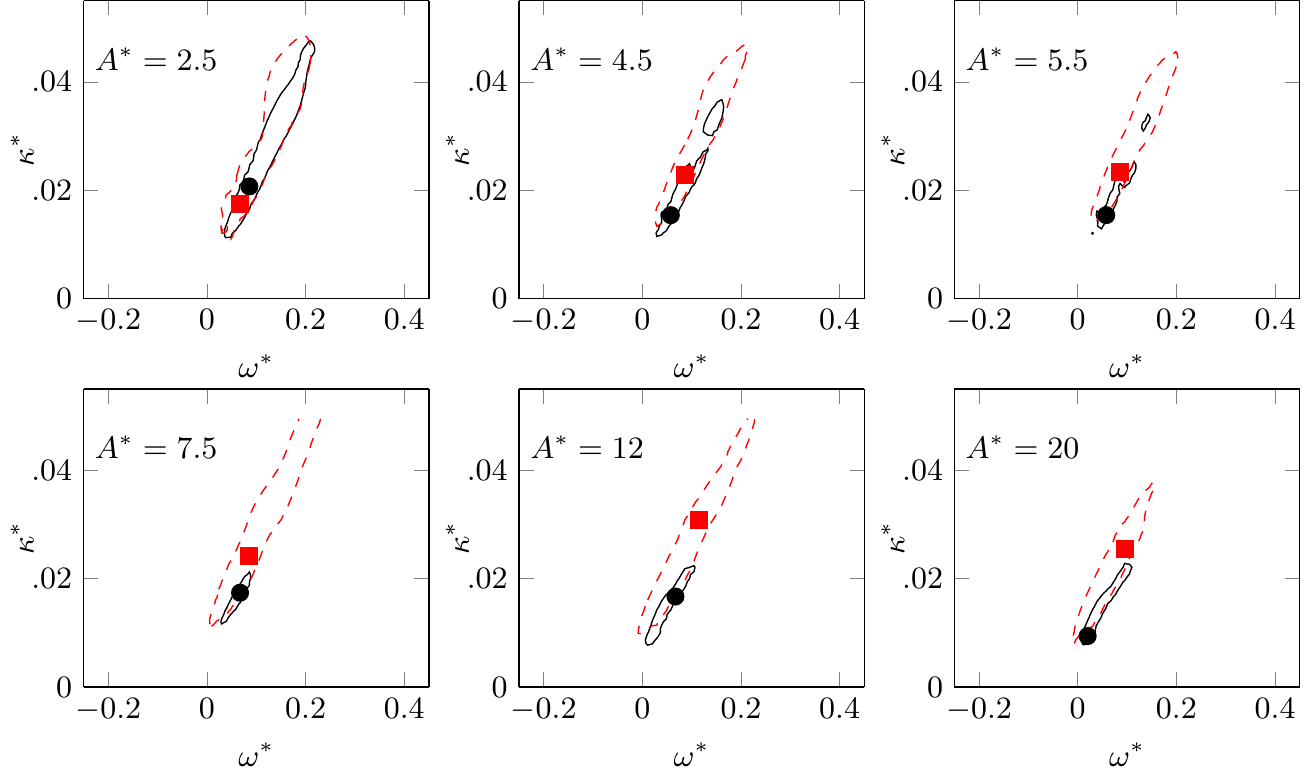}
\caption{Region of large drag reduction, defined as the region where $\R$ is at least $\R_{m,A^*}-0.015$, for various values of $A^*$. Solid black lines: $Re_\tau=200$; dashed red lines: $Re_\tau=1000$. The location of $\R_{m,A^*}$ is marked by a black circle at $Re_\tau=200$ and a red square at $Re_\tau=1000$. 
\label{fig:rmax-shifts}}
\end{figure}

Looking at $\R$ with different scalings leads to important remarks. First, the difference between $\R_{m,A^+}$ and $\R_{m,A^*}$ is small at $A^+=12$. This is explained by the known weak dependence of $\R$ on $A$ at large amplitudes where saturation occurs  \citep{quadrio-ricco-viotti-2009}. Indeed, at lower forcing intensities we observe a stronger drop, which at the lower $Re$ is from $\R_{m,A^+}=0.035 \pm 0.016$ for $A^+=4.5$ to $\R_{m,A^*}=0.29 \pm 0.015$ for $A^*=4.5$. Second, the effect of $Re$ on the value of $\R_{m,A^*}$ is qualitatively similar to the one on $\R_{m,A^+}$, and the shift of the location of the maximum in the $\left( \omega, \kappa \right)$ plane is confirmed. Since the maximum of drag reduction is rather flat, this last observation is better highlighted in figure \ref{fig:rmax-shifts} by observing, at several values of $A^*$, how the shape and size of the region where $\R$ is nearly maximum change while $Re$ increases. This region is identified by the contour line drawn for $\R = \R_{m,A^*} - 0.015$, with 0.015 being a representative value of $\delta \R$. For every value of $A^*$, this region at $Re_\tau=1000$ is found to be larger than the one at $Re_\tau=200$, which is enclosed by the former but for a small part at low wavenumbers and frequencies.

\begin{figure}
\centering
\includegraphics{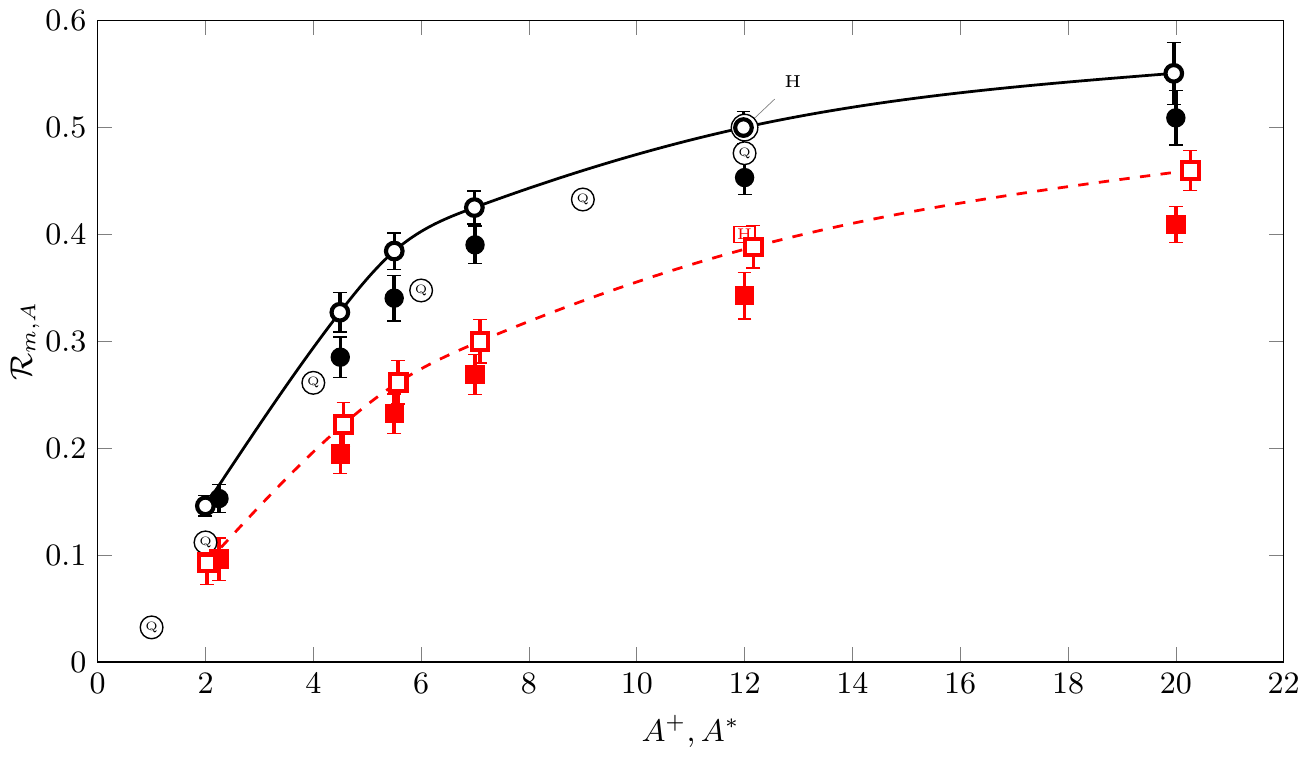}
\caption{Maximum drag reduction $\R_{m,A}$ as function of the forcing amplitude. Black continuous line and symbols are for low-$Re$ data, and red dashed line and symbols are for high-$Re$ data. At each $Re$, open symbols refer to reference scaling ($\R_{m,A^+}$ against $A^+$), and closed symbols to actual scaling ($\R_{m,A^\ast}$ against $A^\ast$). As in figure \ref{fig:OWcomp}, open symbols with letters are literature data. Note that \cite{quadrio-ricco-viotti-2009} assumed reference inner scaling for the optimal $(\omega^+, \kappa^+)$ pair determined at $A^+=12$.}
\label{fig:Rmax}
\end{figure}

A global view of the drag reduction performance of the travelling waves as
determined by the present dataset can be obtained from figure \ref{fig:Rmax}, that shows
for both values of $Re$ how the maximum drag reduction $\R_{m,A}$ increases with the forcing amplitude. It is confirmed that, regardless of the scaling adopted, the general shape of the curve is always that of a saturated growth of $\R$ with $A$, as already observed, for instance, for spanwise wall oscillations by \cite{quadrio-ricco-2004} and for stramwise-travelling waves by \cite{quadrio-ricco-viotti-2009}. The curves with $*$ scaling are consistently lower than the curves with $+$ scaling, and data at $Re_\tau=1000$ are lower than data at $Re_\tau=200$. The
figure also includes a few datapoints from \cite{quadrio-ricco-viotti-2009} at
$Re_\tau=200$. Their point at $A^+=12$ falls slightly below the present data, something that \cite{gatti-quadrio-2013} already documented as an effect of the limited computational domain. However, the larger values of $\R_{m,A^+}$ observed for $A^+ < 12$ in the present study also descend from a better scan of the parameter space. In fact, \cite{quadrio-ricco-viotti-2009} simply assumed a simple (reference) viscous scaling for the location of $\R_{m,A^+}$, whereas here a full parameter study avoids this assumption and determines that $\R_{m,A^+}=0.38 \pm 0.017 $ can still be achieved at $A^+=5.5$ and $Re_\tau=200$, which becomes $\R_{m,A^+}=0.26 \pm 0.020$ at $Re_\tau=1000$. The comparison at $A^+=12$ with the data from \cite{hurst-yang-chung-2014} shows excellent agreement, considered that the slightly higher $\R$ of the high-$Re$ datapoint is due to its lower value of $Re_\tau=800$.

\subsection{Performance indicators}

\begin{figure}
\centering
\includegraphics{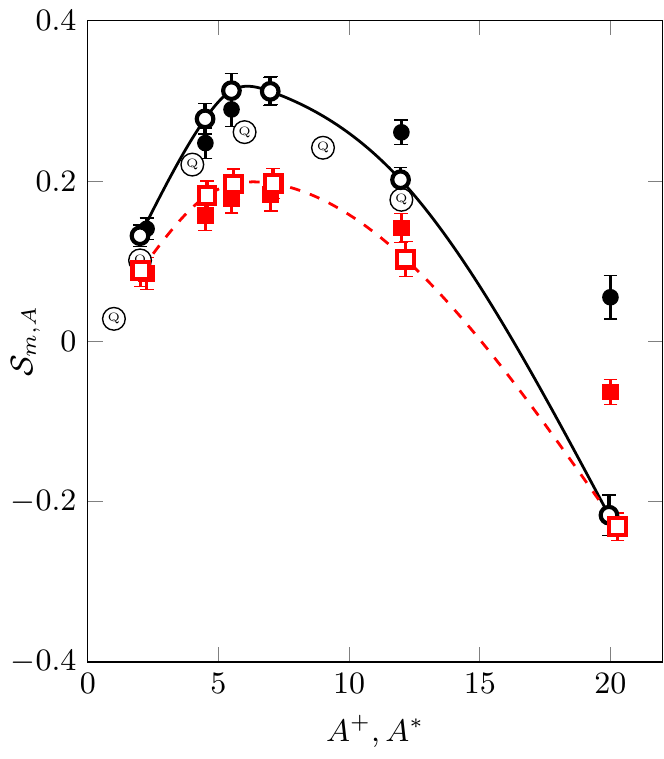}\includegraphics{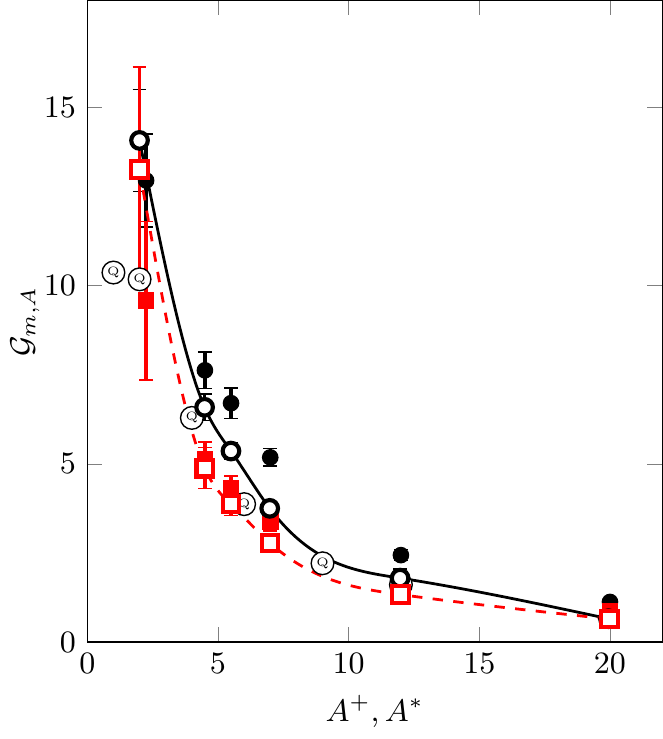}
\caption{Maximum net power saving $\SAV_{m,A}$ (a) and maximum control gain $\G_{m,A}$ (b) as function of the forcing amplitude. Lines and symbols as in figure \ref{fig:Rmax}.
}
\label{fig:SGmax}
\end{figure} 

Active techniques for skin-friction drag reduction should not be characterised by $\R$ alone: the net power saving rate $\SAV$, which also accounts for the energy cost of the control, is an important additional figure of merit for a complete assessment. $\SAV$ can be defined as follows:
\begin{equation}
\SAV = \R - \frac{P_{in}}{P_0} ,
\end{equation}
where $P_0$ is the power (per unit area) required to drive the uncontrolled channel flow, and $P_{in}$ is the control input power, computed here following \cite{baron-quadrio-1996} by neglecting the mechanical losses of a real actuator. 
Figure \ref{fig:SGmax}(a) presents the maximum net power saving $\SAV_{m,A}$ at given amplitude. At both $Re$ the largest $\SAV_{m,A}$ is observed at $A^+ \approx 6$, where the travelling waves yield $\SAV_{m,A^+}=0.31 \pm 0.021$ at $Re_\tau=200$ and $\SAV_{m,A^+}=0.19 \pm 0.018$ at $Re_\tau=1000$. In comparison with existing data from \cite{quadrio-ricco-viotti-2009}, there is a noticeable improvement due to the more detailed scan of the parameter space at low forcing amplitude. As observed by \cite{ricco-quadrio-2008} and \cite{gatti-quadrio-2013}, the power expenditure per unit pumping power $P_{in}/P_0$ decreases with $Re$ at a rate proportional to $Re_\tau^{-0.136}$ and faster than $\R$ in many parts of the control parameter space. This means that $\SAV$ decreases as $\R$, or at a slower rate. $\SAV_{m,A^*}$ is slightly lower than $\SAV_{m,A^+}$ at both $Re$ up to $A^* \approx 7$. In this range of forcing amplitudes the main contributor to $\SAV$ is $\R$, being the input power much smaller, and $\R$ obtained for a given $A^*$ is lower than $\R$ obtained for a given $A^+$. Starting from $A^* \approx 7$, the opposite occurs and $\SAV_{m,A^*}$ becomes larger than $\SAV_{m,A^+}$, because at high forcing amplitude $\SAV$ is dominated by $-P_{in} / P_0$, which is smaller (in absolute value) when forcing is considered at given $A^*$ than at $A^+$.

The control gain $\G$ is defined as the power benefit per unit control power:
\begin{equation}
\G = \frac{\R P_0}{P_{in}} ,
\end{equation}
and is plotted in figure \ref{fig:SGmax} (b). $\G_{m,A^+}$ increases rapidly for decreasing forcing amplitude, since $P_{in}$ is proportional to the square root of $A$ \citep{quadrio-ricco-2011}, while $\R$ varies less than linearly with $A$. At the amplitude value yielding the largest $\SAV_{m,A^+}$ we measure $\G=5.3 \pm 0.24$ at $Re_\tau=200$ and $\G=3.9 \pm 0.31$ at $Re_\tau=1000$, with a mild decrease with $Re$. In fact, the beneficial $Re$ effect on $P_{in}/P_0$ is even more effective on $\G$ than on $\SAV$.

\section{The effect of $Re$ on drag reduction}
\label{sec:re-effect}

We now focus our attention on understanding how the drag reduction performance of the streamwise-travelling waves is affected by an increase of the Reynolds number.

\subsection{Characterizing $\R$ via a power law}

\begin{figure}
\centering
\includegraphics{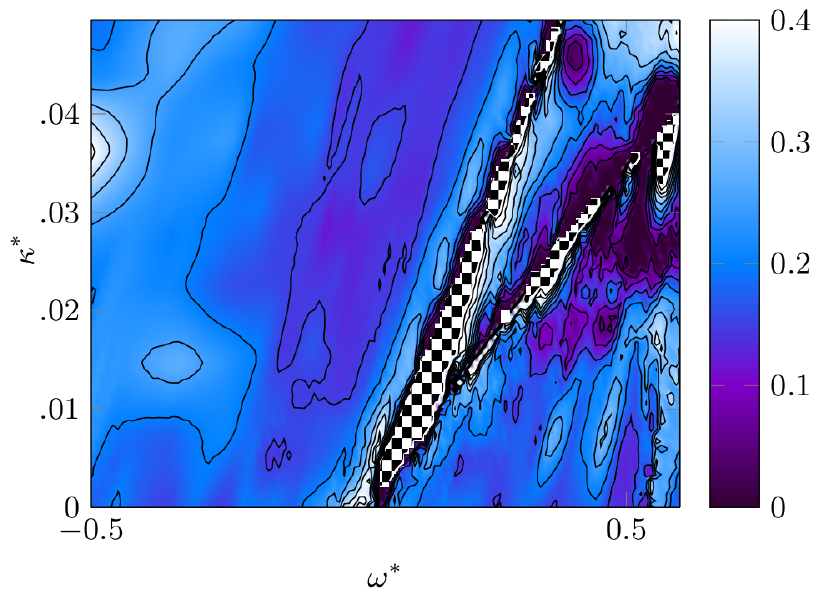}
\caption{Power-law exponent $\gamma$ as a function of $\kappa^*$ and $\omega^*$ at $A^*=12$, measured according to the formula (\ref{eq:gammalog}). Contour lines from 0 to 0.45 in 10 steps. In the regions marked by the chequerboard pattern the exponent $\gamma$ is undefined.}
\label{fig:gamma}
\end{figure}

The effect of $Re$ on $\R$ has been traditionally quantified, as discussed in \S\ref{sec:introduction}, by assuming \citep{choi-xu-sung-2002, moarref-jovanovic-2012, touber-leschziner-2012, duque-etal-2012, belan-quadrio-2013} that $\R$ decreases with $Re$ according to a simple power law, i.e. 
\begin{equation}
\R \propto Re_\tau^{-\gamma} .
\label{eq:gamma}
\end{equation}

The power-law assumption was originally adopted by \cite{choi-xu-sung-2002} to fit early DNS-computed drag reduction results obtained for the spanwise-oscillating wall, and it is not endowed with a specific physical significance. Recently, \cite{gatti-quadrio-2013} and \cite{hurst-yang-chung-2014} observed that $\gamma$ is not a simple constant, but depends on the control parameters $\omega$, $\kappa$ and $A$. The function $\gamma=\gamma(\omega^*, \kappa^*, A^*)$ can easily be constructed from the present dataset. For a given $(\omega^*, \kappa^*, A^*)$ combination, one knows the drag reduction values $\R_{200}$ and $\R_{1000}$ measured at $Re_\tau=200$ and $Re_\tau=1000$, and $\gamma$ can be computed as:
\begin{equation}
  \gamma = \frac{\ln \left( \R_{200} / \R_{1000} \right)}{\ln \left( 200 / 1000 \right)} .
\label{eq:gammalog}
\end{equation}

This quantity is shown in figure \ref{fig:gamma} for $A^*=12$. The details might certainly change depending, for example, on the specific value of $A^*$, on the choice between working at constant $A^+$ or constant $A^*$, or on the choice between using in (\ref{eq:gammalog}) the nominal values of $Re_\tau$ or the actual ones reported in table \ref{tab:discretization-parameters}. However, the emerging picture is consistently as rich as chaotic. The known result that $\gamma$ strongly varies across the parameter space is easily confirmed, and this implies that some regions more than others are sensitive to the increase in $Re$. For example, in most of the ridge of large drag reduction $\gamma$ assumes intermediate values between 0.1 and 0.15, while the largest values are seen in the low-wavenumber region of the drag increase valley.  

The complexity of figure \ref{fig:gamma} graphically illustrates how $\gamma$ is not well suited to comprehensively describe the effect of $Re$ on the drag reduction. Indeed, $\gamma$ is an ill-conditioned quantity, which is sensitive to small uncertainties in $\R$, and according to definition (\ref{eq:gammalog}) it grows to infinity whenever either $\R_{200}$ or $\R_{1000}$ is zero, while it becomes undefined when drag reduction at one $Re$ turns into drag increase at the other $Re$ (see the chequerboard-marked areas in figure \ref{fig:gamma}). However, the most significant conceptual drawback of using $\gamma$ is that, owing to the lack of physical rationale in the assumption of a power-law behaviour, extrapolation of existing data at higher $Re$ (for example those which are typical of aeronautical applications) is entirely arbitrary. In fact, $\gamma$ might well be a function of $Re$ too, as some resulty by \cite{hurst-yang-chung-2014} seem to indicate.  Hence, in our view $\gamma$ should be merely regarded as a (not well-conditioned) indicator with the same limited predictive capability of other, more trivial quantities, like the relative decrease of drag reduction. In the following, we will overcome the problem by introducing a more physically sound indicator.

\subsection{Characterizing $\R$ via the vertical shift $\Delta B$ of the mean velocity profile}
\label{sec:vertical-shift}

When dealing with distributed wall roughness (also including those particular roughness distributions, like streamwise aligned riblets, that reduce friction drag instead of increasing it), it is well known that the consequent friction changes are reflected in the logarithmic region of the velocity profile. According to the classical theory \citep{pope-2000}, the streamwise mean velocity profile $\aver{u}(y)$ over a smooth wall (for which the distinction between the reference inner scaling + and the actual inner scaling * is irrelevant) presents a thin near-wall region, called the viscous sublayer, where $\aver{u}^+ = y^+$, which is connected through the buffer region to the logarithmic layer, where the profile follows a logarithmic law:
\begin{equation}
\aver{u}^+ = \frac{1}{k} \ln y^+ + B .
\label{eq:log-law}
\end{equation} %

In the above expression, $k$ is the von K\'arm\'an constant and $B$ is the so-called additive constant or near-wall intercept. Both constants are known to assume values that weakly depend on $Re$ when $Re$ is low. For instance, \cite{kim-moin-moser-1987} reported $k=0.4$ and $B=5.5$ at $Re_\tau=180$, whereas $k=0.386$ and $B=4.40$ at $Re_\tau=4079$ or $k=0.384$ and $B=4.27$ at $Re_\tau=5200$ have been reported respectively by \cite{bernardini-pirozzoli-orlandi-2014} and \cite{lee-moser-2015}. As stressed by \cite{jimenez-2004}, the additive constant $B$ is determined by the no-slip boundary condition at the wall but, since the logarithmic law is only valid for $y^+ \gg 1$, its value depends on the details of the buffer and viscous layers. In the region further from the wall, the velocity-defect law, which describes the difference between the local mean velocity and the centerline mean velocity $U_c$, takes the following form \citep{pope-2000}:
\begin{equation}
U_c^+ - \aver{u}^+ = -\frac{1}{k} \ln \left( \frac{y}{h} \right) + B_1
\label{eq:defect-law}
\end{equation}
where $B_1$ is a flow-dependent constant representing the difference between the actual centerline velocity $U_c$ and the velocity value obtained by extrapolating the velocity-defect law up to the centerline. In channel flows, $B_1$ is close to zero \citep{pope-2000}. 

If the logarithmic law (\ref{eq:log-law}) and the defect law (\ref{eq:defect-law}) are added together, and $U_c$ is substituted with the more convenient $U_b$ thanks to the relation $U_c^+ = U_b^+ + 1 / \kappa$ \citep{pope-2000}, the use of the definition (\ref{eq:Cf}) for the friction coefficient $C_f$ leads to the following friction law, i.e. an implicit relationship between $C_f$ and the Reynolds number $Re_\tau$:
\begin{equation}
  \sqrt{\frac{2}{C_f}} = \frac{1}{k} \ln Re_\tau + B + B_1 - \frac{1}{k} .
\label{eq:PvK}
\end{equation} 	

When surface roughness is present, a classical statistical description of its effects \citep[see e.g.][]{clauser-1956, jimenez-2004} is via the so-called roughness function, which quantifies the roughness-induced drag change via the downward shift of the velocity profile in the logarithmic layer, i.e. via the (negative) change $\Delta B$ of the additive constant $B$. The same approach is used for the drag-reducing riblets \citep{luchini-manzo-pozzi-1991, garcia-jimenez-2011-a}, which produce a positive $\Delta B$. \cite{luchini-1996} has shown that $\Delta B$ is equivalent to the amount of drag reduction, and is proportional to the so-called protrusion height, defined as the distance between two virtual wall-parallel planes where no-slip boundary conditions for the profiles of the spanwise and streamwise velocity components hold. 

\begin{figure}
\centering
\includegraphics[trim=4pt 0pt 0pt 0pt]{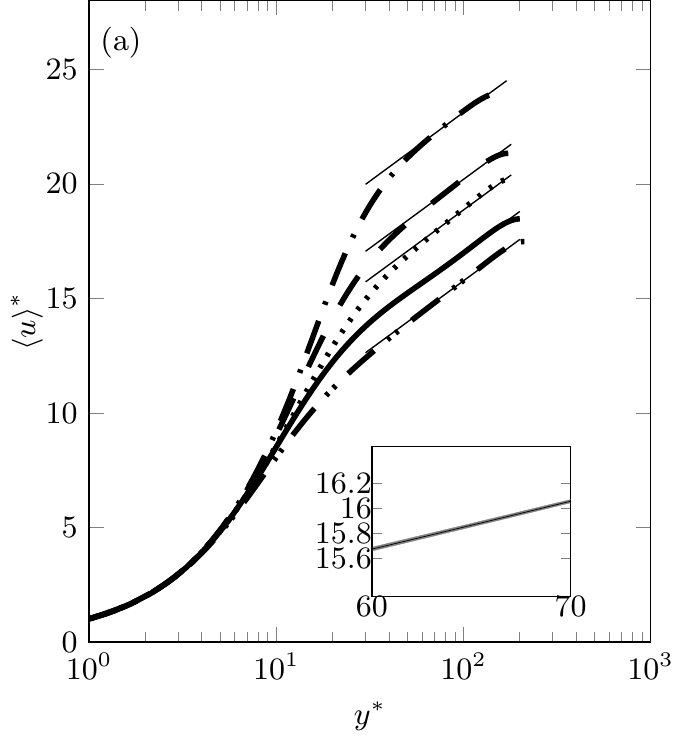}\includegraphics[trim=4pt 0pt 0pt 0pt]{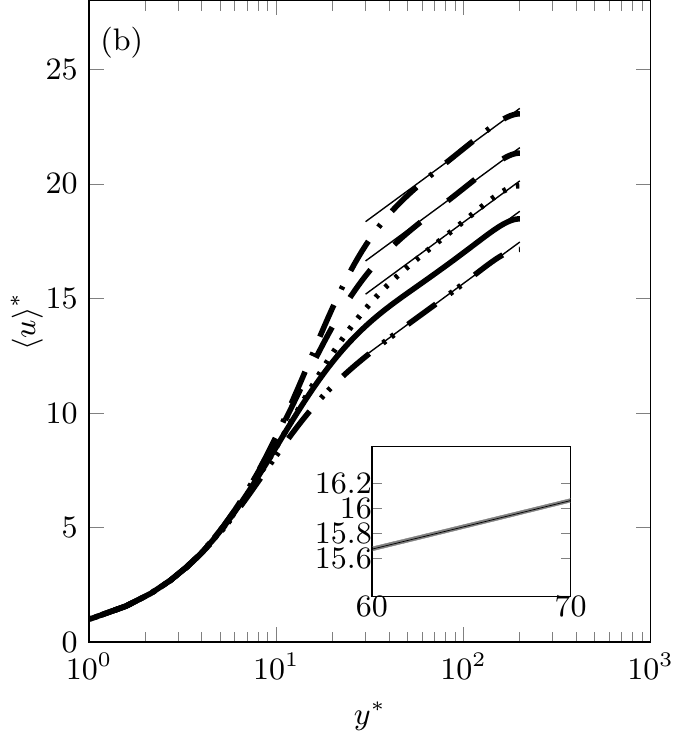}\\
\includegraphics[trim=4pt 0pt 0pt 0pt]{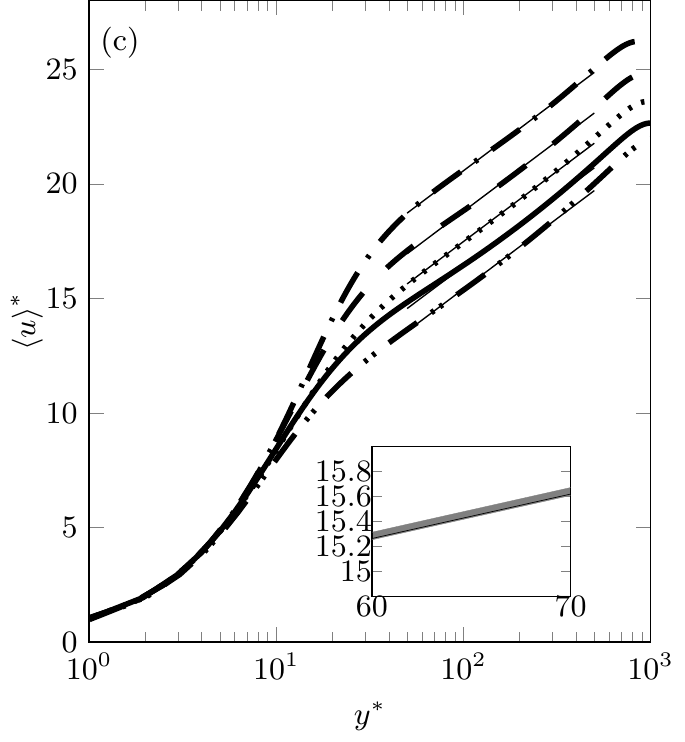}\includegraphics[trim=4pt 0pt 0pt 0pt]{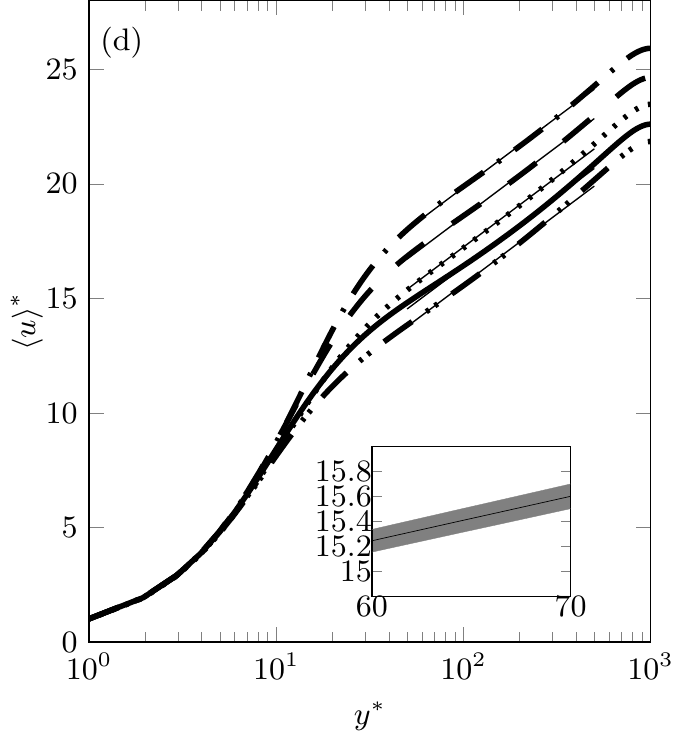}
\caption{Mean velocity profiles obtained from the large-domain simulations reported in the lower half of table \ref{tab:discretization-parameters}. Top: $Re_\tau=200$; bottom: $Re_\tau=1000$. Left: CFR cases; right: CPG cases. The solid line is the reference case and the other lines correspond to control yielding both drag reduction and drag increase (see text). The insets enlarge a portion of the logarithmic layer to show the (very small) statistical uncertainty at 95\% confidence, denoted by the shaded area.}
\label{fig:velprof}
\end{figure}

One is naturally led to think that what applies to changes in wall geometry (roughness) might as well apply to the present, wall-based control strategy. Hence, our initial step consists in verifying whether the travelling waves consistently produce a vertical shift $\Delta B$ of the mean velocity profile. This is a known result \citep[see for instance][]{baron-quadrio-1996,choi-debisschop-clayton-1998,ricco-wu-2004,yudhistira-skote-2011,ricco-etal-2012,touber-leschziner-2012,skote-2014,hurst-yang-chung-2014} that is systematically checked in figure \ref{fig:velprof} (a)-(d). The mean velocity profiles are obtained at both $Re$ from the large-box simulations described in \S\ref{sec:largebox} (discretization details have been reported in the lower half of table \ref{tab:discretization-parameters}). Since we do not want to limit the discussion to the case of maximum drag reduction, at each $Re$ five different cases are computed: the no-control case (solid line) and four additional ones, all with forcing amplitude of $A^+=7$. These are two oscillating wall cases at $T^+=75$ and $T^+=250$, and two travelling waves cases, one with drag reduction at $\omega^+=0.0238$ and $\kappa^+=0.01$, and one with drag increase at $\omega^+=0.12$ and $\kappa^+=0.01$. Every case is computed twice, at the nominal value of $Re_b$ with CFR and at the nominal value of $Re_\tau$ with CPG. For the CPG cases, the viscous quantities mentioned above to define the control conditions must be intended in actual viscous units.

Figure \ref{fig:velprof} (a)-(d) confirms that the spanwise forcing consistently produces a rigid upward (when drag is reduced) or downward (when drag is increased) shift of the mean velocity profile in the logarithmic region, while the parameter $B_1$ and the von K\'arm\'an constant do not change appreciably. In other words, the effect of the spanwise forcing can be quantified via the change $\Delta B$ of the additive constant, regardless of the control type, the value of $Re$, and the simulation strategy. Similar observations were recently put forward by \cite{skote-2014}, who observed changes of $\Delta B$ with drag reduction obtained by spanwise oscillations in a turbulent boundary layer, as well as changes in the von K\'arm\'an constant. The changes in $k$, however, are a direct consequence of the spatial transient present in the boundary layer setting, while in the present parallel flow $\Delta B$ fully describes the effect of the spanwise forcing.

Once the control is seen to modify the mean velocity profile through $\Delta B$, we can follow what has been already done for riblets e.g. by \cite{luchini-manzo-pozzi-1991} and by \cite{garcia-jimenez-2011-a}, and exploit the friction law (\ref{eq:PvK}) to obtain a dimensionless relation between the Reynolds number $Re_\tau$, the drag reduction rate $\R$ and the control-induced change $\Delta B$ of the additive constant. Equation (\ref{eq:PvK}) can be written twice, once for the uncontrolled flow (characterized by $Re_{\tau,0}$ and $C_{f,0}$) and once for the controlled flow (with $Re_\tau$ and $C_f$). By subtracting the latter formula from the former, one obtains:
\begin{equation}
  \sqrt{\frac{2}{C_f}} - \sqrt{\frac{2}{C_{f,0}}} = \frac{1}{k} \ln \frac{Re_\tau}{Re_{\tau,0}} + \Delta B ,
\label{eq:re-effect}
\end{equation}
under the assumption that $k$ and $B_1$ are unaffected by the control. These assumptions are discussed and verified in the Appendix.

Now, if the uncontrolled and controlled flows are compared under the CFR constraint, equation (\ref{eq:re-effect}) can be solved for $\Delta B$ and further manipulated by substituting $C_f = C_{f,0}\left( 1 - \R \right)$ and $Re_\tau = Re_{\tau,0} \sqrt{1-\R}$, to yield:
\begin{equation}
  \Delta B^+ = \sqrt{\frac{2}{C_{f,0}}}\left[ \left( 1 - \R \right)^{-1/2} 
              - 1 \right] - \frac{1}{2 k} \ln \left( 1 - \R \right) .
\label{eq:re-effect-CFR}
\end{equation}

If on the other hand the pressure gradient is kept constant across the comparison (CPG), then by definition $Re_\tau=Re_{\tau,0}$, and the above equation further simplifies to:
\begin{equation}
  \Delta B^* = \sqrt{\frac{2}{C_{f,0}}}\left[ \left( 1- \R \right)^{-1/2} - 1 \right] .
\label{eq:re-effect-CPG}
\end{equation}

It is worth noting that equations (\ref{eq:re-effect-CFR}) and (\ref{eq:re-effect-CPG}) differ from the relationships proposed by \cite{luchini-manzo-pozzi-1991} and by \cite{garcia-jimenez-2011-a} for the turbulent flow over a surface with riblets. In fact, those are linearized under the assumption of small $\R$, which is definitely a reasonable assumption for riblets but does not apply to the present forcing, which is capable to yield quite large $\R$ and consequently is better described by the fully non-linear expression. 

\begin{figure}
\centering
\includegraphics{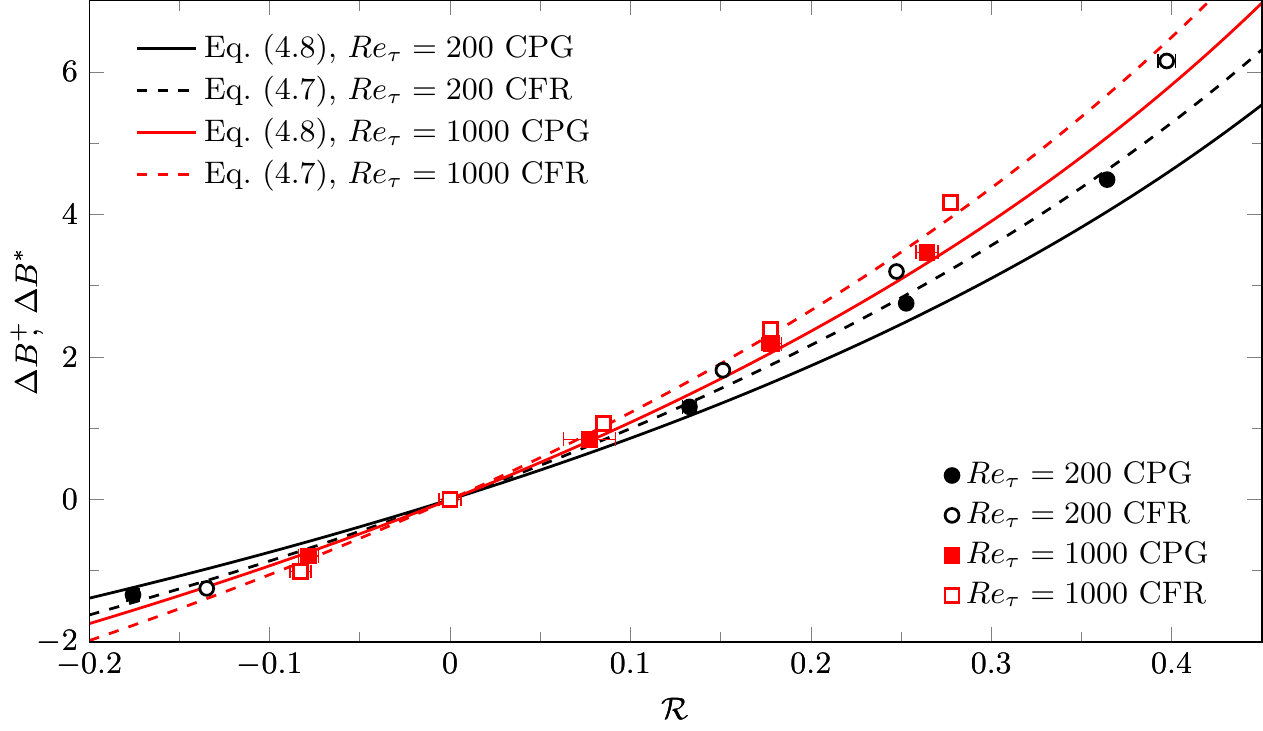}
\caption{Relationship between the upward shift $\Delta B$ and the drag reduction rate $\R$. Symbols are present large-box DNS data, and lines are predictions from (\ref{eq:re-effect-CFR}) and (\ref{eq:re-effect-CPG}) for the CFR and CPG cases.}
\label{fig:R-deltaB}
\end{figure}

The main point is that, since $C_{f,0}$ is a function of $Re_b$ only, equations (\ref{eq:re-effect-CFR}) and (\ref{eq:re-effect-CPG}) link $\Delta B$ and $\R$, provided $Re_b$ is large enough for the friction law (\ref{eq:PvK}) to hold. This can easily be tested with our DNS data, as shown in figure \ref{fig:R-deltaB}, where data from the large-box simulations already considered in figure \ref{fig:velprof} are used. The uncertainty on $\Delta B$ has been computed at 95\% confidence level through partial-derivative-based Gaussian error propagation of the uncertainties on $C_f$ and $C_{f,0}$, similarly to what has been done for $\R$ in \S\ \ref{sec:uncert}, but using the definitions of $\Delta B$ given in equations (\ref{eq:re-effect-CFR}) and (\ref{eq:re-effect-CPG}). At $Re_\tau=1000$ the agreement is excellent, for both the CPG and CFR cases. The agreement is less good at $Re_\tau=200$, and it is interesting to note that the largest deviation occurs for the CFR case with largest $\R$ at $Re_\tau=200$, i.e. for the case possessing the lowest actual $Re_\tau$ (namely $Re_\tau=173$). It comes at no surprise that, at such low $Re_\tau$, the assumptions underlying the friction law (\ref{eq:PvK}), which implies that the flow rate is well approximated by the wall-normal integral of the logarithmic velocity profile, do not apply in full. 

\begin{figure}
\centering
\includegraphics{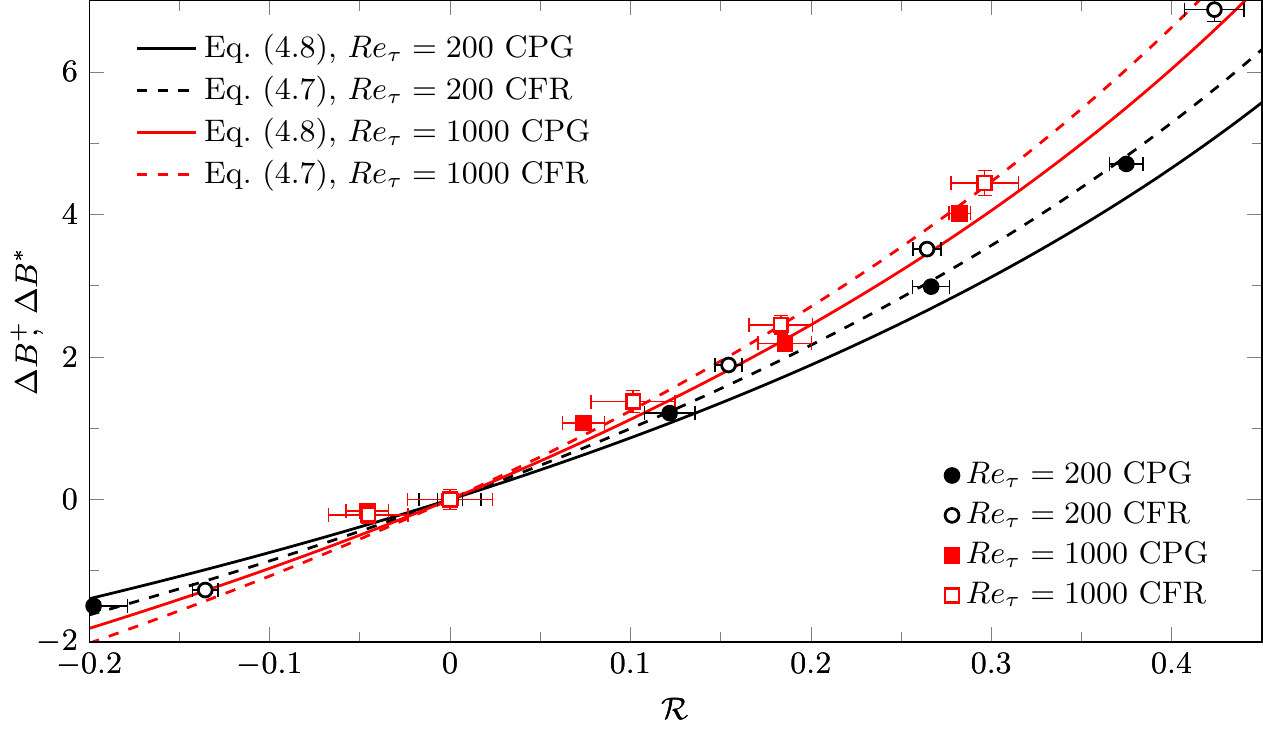}
\caption{Relationship between the upward shift $\Delta B$ and the drag reduction rate $\R$. Symbols are present small-box DNS data, and lines are predictions from (\ref{eq:re-effect-CFR}) and (\ref{eq:re-effect-CPG}) for the CFR and CPG cases.}
\label{fig:R-deltaB-SMALL}
\end{figure}

Quite surprisingly, a very similar picture is observed in figure \ref{fig:R-deltaB-SMALL} for the small-box simulations, where the mean velocity profile in the outer region is expected to be affected by the size of the computational domain. Of course, the five CPG datapoints are not contained in Tab. \ref{tab:discretization-parameters}, which lists only CFR cases for the small-box dataset, and have been produced on purpose. Despite the larger statistical uncertainty of $\R$ and the indeed quite small extent of healthy logarithmic region, which translates into a non-negligible uncertainty for $\Delta B$, the data at $Re_\tau=1000$ agree fairly well with equations (\ref{eq:re-effect-CFR}) and (\ref{eq:re-effect-CPG}), while those at $Re_\tau=200$ do not, because $Re$ is too low, in analogy with the corresponding data from the large-box simulations.

\begin{figure}
\centering
\includegraphics{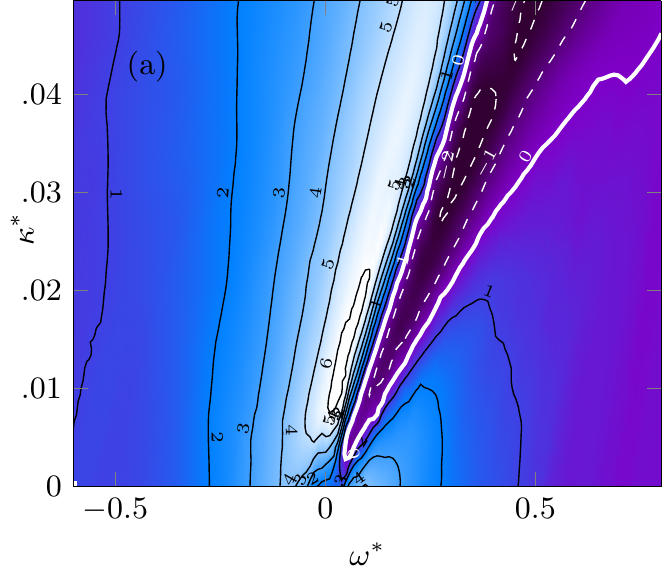}\includegraphics{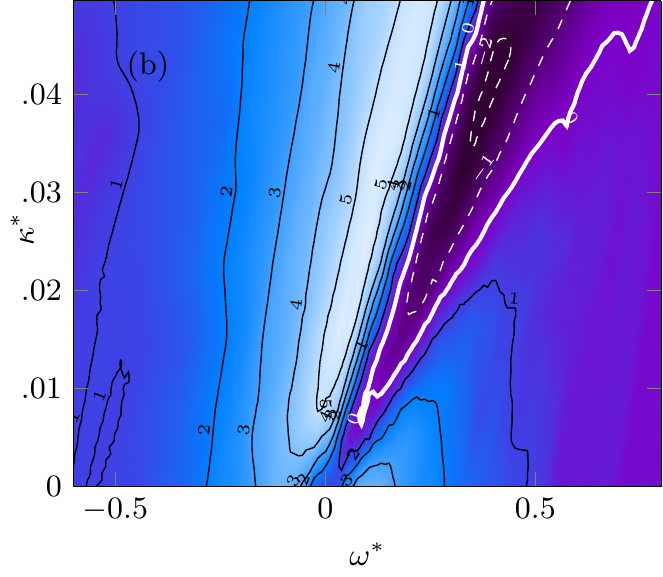}
\caption{Map of $\Delta B^*$ for streamwise travelling waves at $A^*=12$, for $Re_\tau=200$ (a) and $Re_\tau=1000$ (b). Contours are spaced by 1, negative contours are dashed. The thick white line corresponds to $\Delta B^* = 0$.}
\label{fig:deltaB}
\end{figure}

The evidence that equations (\ref{eq:re-effect-CFR}) and (\ref{eq:re-effect-CPG}) describe well the relationship among $C_{f,0}$ (hence $Re_\tau$), $\R$ and $\Delta B$ at sufficiently high $Re$ implies that these equations also describe the effect of $Re$ on $\R$, provided information is known on the $Re$-dependency of $\Delta B$. This information can be educed from the small-box database introduced in \S\ref{sec:smallbox}. Figure \ref{fig:deltaB} shows the map of $\Delta B^*$ at $Re_\tau=200$ and $Re_\tau=1000$, in actual viscous scaling at $A^*=12$. Instead of measuring $\Delta B$ from the velocity profiles, $\Delta B$ is computed here {\em ex-post} from the independently validated values of $\R$ through equation (\ref{eq:re-effect-CFR}). In this way, the computed value of $\Delta B$ matches the measured value at $Re_\tau=1000$, while it is consistently slightly smaller  at $Re_\tau=200$ (see figures \ref{fig:R-deltaB} and \ref{fig:R-deltaB-SMALL}). This estimate of $\Delta B$ at low $Re$ is still meaningful, as it can be interpreted to represent the $\Delta B$ value epurated from the low-$Re$ effects that are due to deviations from the predictions of the friction law. 

\begin{figure}
\centering
\includegraphics{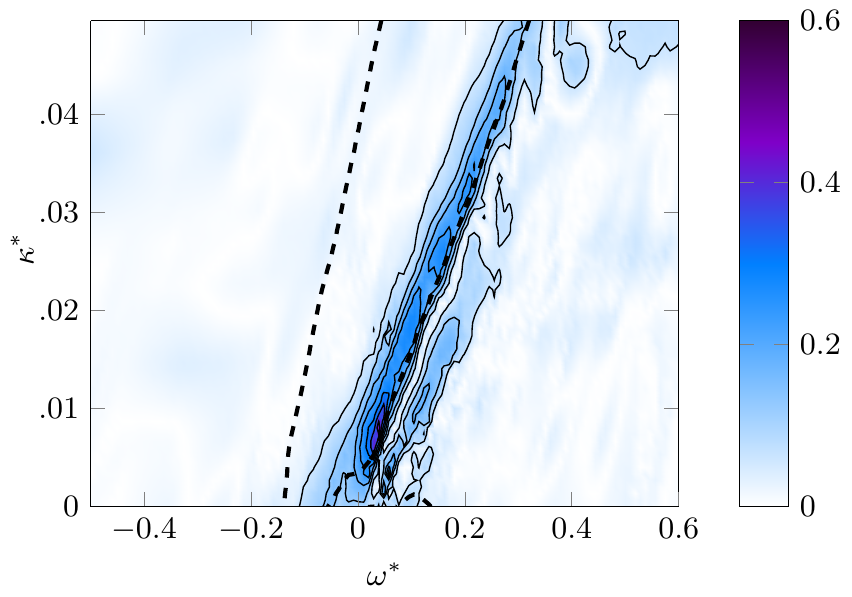}
\caption{Map of the difference between $\Delta B^*$ at $Re_\tau=200$ and $\Delta B^*$ at $Re_\tau=1000$ for $A^*=12$, relative to the reference value $B=5.2$, obtained for the uncontrolled flow at $Re_\tau=200$. Contour lines from 0.06 to 0.6 in 10 steps. The thick dashed line marks the region of large drag reduction, i.e. $\R>0.3$, at $Re_\tau=200$.}
\label{fig:deltadeltaB}
\end{figure}

When compared to the two maps in figure \ref{fig:r-A12}, where $\R$ is observed to decrease significantly across the whole plane, the two maps of $\Delta B^*$ appear much more similar, if exception is made for a localized change near the origin of the $(\omega^*, \kappa^*)$ plane. Figure \ref{fig:deltadeltaB} emphasizes the relative changes of $\Delta B^*$ with $Re$, by plotting the difference between the two previous maps, i.e. the pointwise difference between $\Delta B^*$ at $Re_\tau=200$ and $\Delta B^*$ at $Re_\tau=1000$, normalized by the reference value $B=5.2$ of the uncontrolled flow at $Re_\tau=200$. The figure strengthens the previous claim that $\Delta B$ is roughly constant with $Re$, once low-$Re$ effects are discarded. In fact, except for a thin stripe located on the right of the drag reduction ridge, the variation of $\Delta B^*$ with $Re$ is essentially zero. In this region, which includes the whole area of large drag reduction except its right boundary, the function $\Delta B^* (\omega^*, \kappa^*)$ is solely determined by the control parameters, and equation (\ref{eq:re-effect}) can be used to describe the decrease of $\R$ with $Re$. The thin stripe, on the other hand, is a low-$Re$ effect that represents a non-trivial feature specific to the present control technique. 

Although this is not enough for the ultimate demonstration that $\Delta B^*$ does not depend on $Re$, the scenario depicted above is more than reasonable, and is reinforced by the parallel made with riblets. Moreover, additional data exist to support it. For example, \cite{hurst-yang-chung-2014} plot in their figure 16 mean streamwise velocity profiles for channel flow modified by stationary waves at $\omega^+=0$ and $\kappa^+=0.008$ with $A^+=12$ and $Re_\tau = \left\{200, 400, 800, 1600 \right\}$. By comparison with the reference profiles, they report $\Delta B^+ = \left\{7.6, 6.2, 5.5, 5.0 \right\}$ respectively. The changes clearly are progressively decreasing as $Re$ is increased. Moreover, it should be noted that the point at $\omega^+=0$ and $\kappa^+=0.008$ lies in the small area where the residual $Re$-effect is maximum (see our figure \ref{fig:deltadeltaB}) and some change $\Delta B$ is expected. Lastly, \cite{hurst-yang-chung-2014} adopted a reference viscous scaling, which by itself induces a change in $\Delta B$: as $Re$ increases, $\R$ shrinks, $u_\tau / u_{\tau,0}$ increases and the actual forcing amplitude $A^*$ decreases. Hence, data from \cite{hurst-yang-chung-2014} are compatible with the present work in suggesting that for a large enough $Re$ the function $\Delta B(\omega^*, \kappa^*)$ characterizes the forcing, and becomes constant with $Re$ above a threshold value. We estimate this minimum $Re$ to be $Re_\tau \approx 2000$.

\section{Concluding discussion}

In this study a large drag reduction DNS database has been produced for a turbulent plane channel flow subject to a spanwise forcing. 4020 simulations have been used to describe how increasing the value of the Reynolds number from $Re_\tau=200$ to $Re_\tau=1000$ affects drag reduction, and to propose a rationale behind the observed performance deterioration. To the authors' knowledge, this is the first study on spanwise forcing that includes a wide range of forcing amplitudes, as well as Contant Pressure Gradient (CPG) data at different values of $Re$. The large size of the numerical study has been possible thanks to the use of relatively small computational domains, which at the same time also constitutes its main limitation. This strategy has been already proved by \cite{gatti-quadrio-2013} to successfully provide useful and accurate information about changes in friction drag. Moreover, the numerical results and the main conclusions are corroborated by 20 additional cases where a large computational domain is employed. The main findings of this study can summarized as follows.

\begin{enumerate}

\item The existing information regarding spanwise forcing has been significantly extended. Thanks to the depth of the numerical study, the maximum {\em net} saving rate of streamwise-travelling waves at $Re_\tau=200$ is $0.31 \pm 0.021$, condition at which they produce a drag reduction of $0.382 \pm 0.002$ and a gain of $5.3 \pm 0.24$; these figures become $0.19 \pm 0.002$, $0.255 \pm 0.02$ and $3.9 \pm 0.31$ at $Re_\tau=1000$. We have shown how the scaling choices (viscous units based on the reference friction velocity, i.e. $``+"$ scaling, or viscous units based on the actual friction velocity of the drag-reduced flow, i.e. $``*"$ scaling) impact the results, and the ensuing differences have been discussed. Though less practical to implement, actual $*$ scaling has been shown to be more advantageous to describe phenomena and quantities that are expected to scale in viscous units, such as the shape of the mean velocity profile in the viscous sublayer or the shape of skin-friction drag reduction maps at different $Re$.

\item The deterioration of performance with $Re$ has been confirmed, in agreement with the majority of existing data. The often assumed but arbitrary power-law decrease of drag reduction with $Re$, i.e. $\R \sim Re_\tau^{-\gamma}$, yields values of the exponent $\gamma$ which vary strongly across the space of control parameters. The lack of rationale of a power-law dependency is such that a $\gamma$-based extrapolation at higher $Re$ of existing data should be regarded as potentially unreliable. 

\item The classic argument linking the skin-friction drag changes of a rough wall to the vertical shift $\Delta B$ of the logarithmic portion of the mean velocity profile has been shown to apply to the case of spanwise forcing. A non-linear expression has been derived that can be specialized to the CFR or CPG cases. As for drag-reducing riblets, characterizing drag reduction through $\Delta B$ is more informative than the simple statement of "percentage drag reduction". Indeed, $\Delta B$ already contains the necessary $Re$ dependency through the friction law and, if the control parameters in actual units are constant, it does not significantly change once $Re$ is above the threshold where the mean velocity profile features a well-defined logarithmic region, and the validity of the friction law (\ref{eq:PvK}) is ensured. 

\begin{figure}
\centering
\includegraphics[trim=4pt 0pt 0pt 0pt]{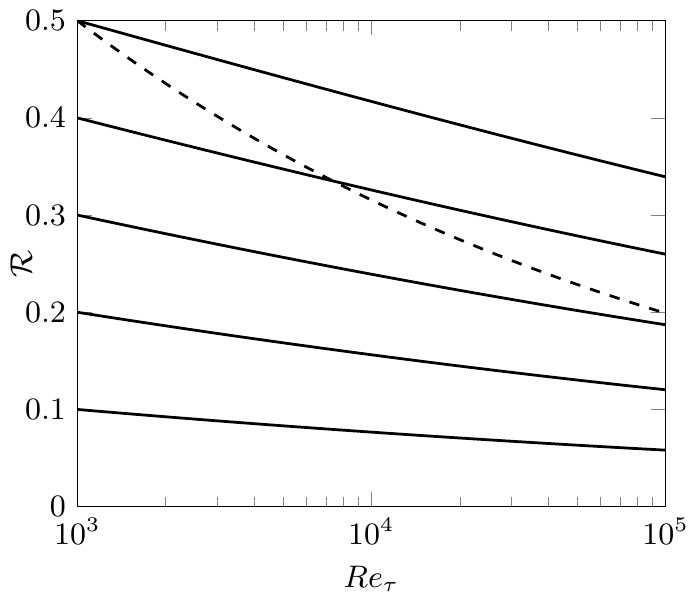}\includegraphics[trim=5pt 0pt 0pt 0pt]{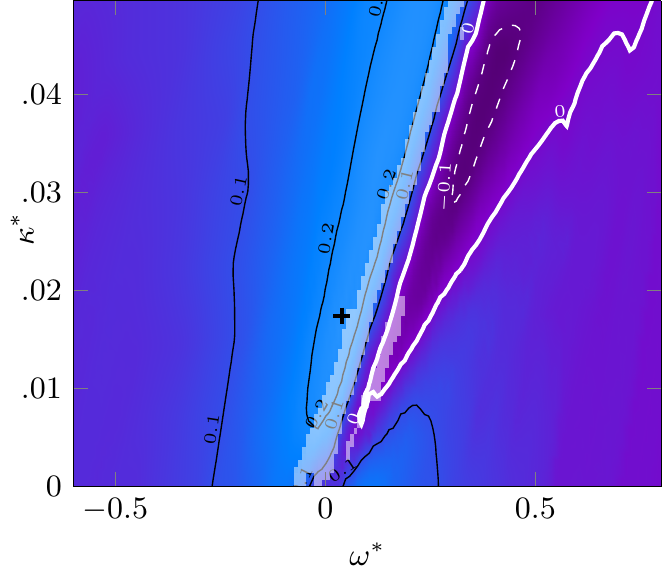}
\caption{Extrapolation of drag reduction data from $Re_\tau=1000$ to $Re_\tau=100,000$. (a) Continuous lines are prediction by equation (\ref{eq:re-effect}), while the dashed line shows a decrease of $\R \propto Re_\tau^{-0.2}$. (b) Extrapolated drag reduction map at $Re_\tau=10^5$ for $A^*=12$, lines and symbols as in figure \ref{fig:R-A12}. The white shaded area is excluded from the contour plot as there $\Delta B$ is found to vary between $Re_\tau=200$ and $Re_\tau=1000$.}
\label{fig:extrap}
\end{figure}

\item Under the assumption that $\Delta B^*$ measured in the present work at $Re_\tau=1000$ is already $Re$-independent, equation (\ref{eq:re-effect}) can be used to extrapolate drag reduction at higher $Re_\tau$, as done in figure \ref{fig:extrap}(a), showing that a drag reduction of $\R=0.5$ at $Re_\tau=1000$ translates into $\R=0.34$ at $Re_\tau=10^5$. The decrease is still significant but not as dramatic as the low-$Re$ evidence suggests: this can be easily appreciated by looking at the dashed line, that describes $\R \sim Re_\tau^{- \gamma}$ with $\gamma=0.20$. The whole drag reduction map at $A^*=12$ is extrapolated from $Re_\tau=1000$ to $Re_\tau=10^5$ in figure \ref{fig:extrap}(b). The maximum drag reduction in the plane reduces from 0.34 to 0.23, which is still a sizeable amount considered the two-decades increase of $Re$. 

\item While we are not in the position to make a claim of generality, we believe that the line of reasoning developed in this paper for spanwise forcing might apply to other wall-based active control techniques, as long as they do not disrupt the shape of the mean velocity profile. The relation among $\Delta B$, $C_{f,0}$ and $\R$ and the observation that $\Delta B$ becomes $Re$-independent provide an unbiased framework to describe the effect of the Reynolds number on turbulent drag reduction obtained via wall-based control. For riblets, $\Delta B$ depends on the riblet geometry alone. For spanwise forcing, a given set of control parameters determines the properties of the generalized Stokes layer \citep{quadrio-ricco-2011}, whose features, such as its thickness \citep{quadrio-ricco-2011} or spanwise shear stress at a certain wall distance \citep{yakeno-hasegawa-kasagi-2014}, determine the shift $\Delta B$ or, equivalently, the $Re$-dependent drag reduction rate $\R$ according to equations (\ref{eq:re-effect-CFR}) or (\ref{eq:re-effect-CPG}). Hence, the comparison between wall-based control at different Reynolds numbers should be made in terms of $\Delta B$, not of $\R$.

\item Although the present results are sound from a quantitative standpoint, we believe that an accurate DNS study of the drag-reduction properties of the travelling waves at higher $Re$ is still needed. Such ultimate study would be forcedly less vast than the present one, but could be designed by leveraging  the available information. The view presented in this paper, with its Achilles' heel of being based on small-box simulations, will surely benefit from an independent confirmation. Moreover, $Re_\tau=1000$ is still smaller than the minimal $Re$ at which the values of $\Delta B$ are expected to become fully $Re$-independent. We stress that this minimal $Re$, suggested to be $Re_\tau \approx 2000$, is quite smaller than the minimal $Re$ required for the mean velocity profile to show an incipient true logarithmic region, observed for example by \cite{lozanoduran-jimenez-2014b} at $Re_\tau=4200$ and by \cite{lee-moser-2015} at $Re_\tau=5200$. Anyway, this ultimate study is all what is left to do for the complete characterization of the drag reduction capabilities of the travelling waves, at any (high enough) $Re$.

\end{enumerate}

\vspace{0.25cm}

Although so far we have deliberately avoided to mention turbulent structures, by strictly limiting our discussion to the simplest first-order statistics, we would like to close this paper with a remark of a more general nature.
 
On a fundamental level, the dependence of $C_f$ upon $Re$ may be seen to reflect the variable extension of the different portions of a turbulent wall flow, in particular the inner and the outer layers, and the way they interact with each other. The importance of the inner layer has been probably overemphasized in the last decades: indeed, at the low values of $Re$ typical of most studies addressing near-wall turbulence and its control, the inner layer occupies a significant portion of the whole flow.  At higher $Re$, however, the wall-normal extension of the inner layer becomes progressively smaller with respect to the outer layer, and at application-level $Re$ most of the boundary layer is just outer layer. It is also known, thanks for example to the FIK identity \citep{fukagata-iwamoto-kasagi-2002}, that the portion of Reynolds shear stresses located in the outer layer carries an increasing contribution to the skin-friction coefficient as $Re$ increases. These Reynolds stresses are due to the very large turbulent structures \citep{ganapathisubramani-longmire-marusic-2003, guala-hommema-adrian-2006} that populate the outer layer and modulate the smaller structures residing near the wall \citep{brown-thomas-1977,ganapathisubramani-etal-2012,lozanoduran-jimenez-2014}. This modulation has been confirmed also in the case of drag-reducing flows \citep[see for example][]{touber-leschziner-2012}. 

Those among the present results which are computed in small computational domains only partially represent the large structures of the flow. In fact, only those structures whose width fits into the domain are actually well resolved \citep{flores-jimenez-2010}, with their longer streamwise extent ending up as indefinitely long owing to the periodic boundary conditions. Even though physically incorrect, these large scales still interact  healthily with the well-resolved smaller scales, as pointed out by \cite{lozanoduran-jimenez-2014b}, who observed that small computational boxes with spanwise width larger than $L_z /h = \pi / 2$ still yield correct low-order statistics. However, the present values of $L_z$ are well below this threshold, so that a significant part of the large scales, that at $Re_\tau=1000$ already carry a non-negligible contribution to Reynolds stresses and hence to turbulent drag \citep{ganapathisubramani-longmire-marusic-2003}, are not resolved. This explains the significant deviations of the values of $C_f$ from the correct value observed for the uncontrolled flow. Nonetheless, the nearly correct prediction of changes in $C_f$, i.e. drag reduction, via small-box simulations suggests that the large outer structures might not significantly interfere with the working mechanism of wall-based strategies for drag reduction. The control action takes place at the wall and targets turbulence in the very vicinity of the wall, and the largest eddies which attach to the wall \citep{lozanoduran-jimenez-2014} simply perceive a modified inner layer with a lower mean shear. In other words, the control performance deterioration occurs mainly because of the geometric effect of the shrinking near-wall region, which contributes progressively less to turbulent drag. The large structures, which contribute to turbulent friction via the Reynolds shear stress produced in the outer region, do not directly jeopardize the drag-reducing action of the wall-based spanwise forcing. This does not preclude that, should a control technique succeed in targeting the outer structures directly, additional turbulent drag reduction at high $Re$ could be obtained. The present results thus blend nicely with the result, presented by \cite{iwamoto-etal-2005}, that artificially removing the turbulent fluctuations in a thin wall layer of constant thickness when measured in inner units produces a reduction of friction which is unexpectedly robust with respect to the increase in $Re$. There is thus no need to control the outer structures directly to achieve drag reduction at high $Re$, although of course being able to do so would yield additional benefits.

\section*{Acknowledgments}
Preliminary versions of this work have been presented by MQ at the ERCOFTAC Flow Control and Drag Reduction meeting in Cambridge, April 2015, and by DG at the XV Euromech Turbulence Conference in Delft, August 2015. Computing time has been provided by the Italian Supercomputing Center CINECA (through a LISA 2013 grant), and by the computational resource ForHLR Phase I funded by the Ministry of Science, Research and the Arts Baden-W\"urttemberg and DFG ("Deutsche Forschungsgemeinschaft"). Sergei Chernyshenko is thanked for an interesting discussion on the topic in Lyon during ETC 2013.

\section*{Appendix} 

In \S\ref{sec:vertical-shift} equations (\ref{eq:re-effect-CFR}) and (\ref{eq:re-effect-CPG}) reciprocally linking drag reduction, the friction coefficient and $Re$ have been deduced under the assumption that the von K\'arm\'an constant $k$ is independent upon $Re$, and the intercept $B_1$ is negligible for channel flows. The validity of these assumptions can be verified by measuring $k$, $B$ and $B_1$ for the whole set of 20 large-box simulations by fitting the log-law (\ref{eq:defect-law}) to the mean velocity profiles shown in figure \ref{fig:velprof}. The results are summarized in Table \ref{tab:log-fit}. 

The extrema of the least-squares fitted region in wall-normal direction are chosen in such a way that the width of the fitted region is as large as possible, while keeping a correlation coefficient $r^2 \geq 0.9995$ (where $r^2=1$ means perfect fit). The same procedure was employed by \cite{lee-moser-2015} with $r^2 = 0.9999$ at $Re_\tau = 5186$. The present, slightly lower threshold for $r^2$ descends from the lower $Re$, which implies a shorter logarithmic region. The lower bound of the fit is set at least ten wall units past the relative minimum of the diagnostic function $y^+ \mathrm{d}\aver{u}^+ / \mathrm{d}y^+$. Note that choosing the range for this curve fit is somewhat arbitrary \citep{lee-moser-2015}.

The near-wall intercept $B$ is a function of $k$ and is very sensitive to small variation thereof. Thus, small uncertainties on $k$ due to the arbitrariness of the log-region range strongly affect $B$ and make its measure not readily comparable among different simulations. Therefore, the numerical value of $k$ in the controlled cases is kept equal to the uncontrolled value while the validity of the aforementioned assumption is verified indirectly by monitoring that the change in $k$ is smaller than 0.01. In fact, the value of $k$ is known to weakly influence the way the logarithmic and defect-laws (\ref{eq:log-law}) and (\ref{eq:defect-law}) approximate the mean velocity profile \citep{delalamo-etal-2004}.

\begin{table}
\begin{tabular*}{1.0\textwidth}{@{\extracolsep{\fill}} l l r r r r r r r}

\multicolumn{9}{c}{Full-size simulations at Constant Flow Rate (CFR)} \\

\hline

$Re_\tau$  &  $Re_b$  &  $\kappa^+$  &  $\omega^+$  &  $k$      &  $B^*$    &  $B_1^*$  &  $\R$  &  $\delta \R$ \\[10pt]

199.9      &  6361    &  -              &  -         &  0.390 &  5.151  &  -0.29   &  -      & 0.003       \\ 
173.4      &  6361    &  0              &  0.0838    &  0.390 &  8.351  &  -0.32   &  0.248  & 0.003       \\
184.1      &  6361    &  0              &  0.0251    &  0.386 &  6.965  &  -0.30   &  0.151  & 0.003       \\
155.2      &  6361    &  0.01           &  0.0238    &  0.390 & 11.305  &  -0.32   &  0.397  & 0.004       \\
213.0      &  6361    &  0.01           &  0.12      &  0.386 &  3.904  &  -0.29   & -0.135  & 0.006       \\

\hline

998.7      &  39990   &  -              &  -         &  0.386 &  4.494  &  0.26   &  -       & 0.003      \\ 
905.7      &  39990   &  0              &  0.0838    &  0.386 &  6.885  &  0.26   &  0.178   & 0.002      \\
955.2      &  39990   &  0              &  0.0251    &  0.378 &  5.560  &  0.25   &  0.085   & 0.003      \\
849.0      &  39990   &  0.01           &  0.0238    &  0.386 &  8.665  &  0.10   &  0.277   & 0.005      \\
1039.2     &  39990   &  0.01           &  0.12      &  0.385 &  3.484  &  0.46   & -0.083   & 0.003      \\

\multicolumn{9}{c}{} \\

\multicolumn{9}{c}{Full-size simulations at Constant Pressure Gradient (CPG)} \\

\hline

$Re_\tau$  &  $Re_b$  &  $\kappa^*$  &  $\omega^*$  &  $k$      &  $B^*$    &  $B_1^*$  &  $\R$  &  $\delta \R$ \\[10pt]

200.0        &  6358    &  -              &  -         &  0.390 &  5.187  &  -0.30   &  -      & 0.003       \\ 
200.0        &  7356    &  0              &  0.0838    &  0.387 &  7.940  &  -0.31   &  0.253  & 0.002       \\
200.0        &  6827    &  0              &  0.0251    &  0.381 &  6.487  &  -0.31   &  0.133  & 0.004       \\
200.0        &  7974    &  0.01           &  0.0238    &  0.390 &  9.675  &  -0.30   &  0.364  & 0.002       \\
200.0        &  5864    &  0.01           &  0.12      &  0.388 &  3.849  &  -0.32   & -0.176  & 0.003       \\

\hline

1000.0       &  19995   &  -              &  -         &  0.387 &  4.525  &  0.25   &  -       & 0.006      \\ 
1000.0       &  44119   &  0              &  0.0838    &  0.386 &  7.711  &  0.10   &  0.178   & 0.005      \\
1000.0       &  41635   &  0              &  0.0251    &  0.380 &  5.371  &  0.27   &  0.077   & 0.001      \\
1000.0       &  46627   &  0.01           &  0.0238    &  0.388 &  7.993  &  0.09   &  0.264   & 0.006      \\
1000.0       &  38514   &  0.01           &  0.12      &  0.390 &  3.735  &  0.28   & -0.078   & 0.006      \\

\end{tabular*}
\caption{Values of the von K\'arm\'an constant $k$, the intercepts $B^*$ and $B_1^*$, the drag reduction rate $\R$ and its uncertainty $\delta \R$, for the 20 large-box cases detailed in table \ref{tab:discretization-parameters}, identified with their control parameters and their actual values of $Re$.}
\label{tab:log-fit}
\end{table}

\end{document}